\documentclass{article}
\usepackage[pdftex]{color}
\usepackage{pdfcolmk}
\usepackage[pdftex]{graphicx}
\usepackage{epstopdf}
\usepackage{mathptmx}
\usepackage{natbib}
\usepackage{authblk}
\usepackage[a4paper,head=17mm,headsep=7mm,top=0.8in,left=0.9in,right=0.9in,bottom=0.8in]{geometry}

\usepackage{amsmath}
\usepackage{amsfonts}
\usepackage{amssymb}
\usepackage{amsbsy}

\newcommand{\E}[1]{\mathbb{E}\left[\, #1 \,\right]}

\newcommand {\ento}{\frac{1}{2}}
\newcommand {\lp}{\left(}
\newcommand {\rp}{\right)}

\newcommand{\gradgzfj}{\bmG_j}
\newcommand{\gradgzf}{\bmG}

\newcommand{\rma}{\mathrm{a}}

\newcommand{\rmd}{\mathrm{d}}

\newcommand{\rmf}{\mathrm{f}}

\newcommand{\rmT}{\mathrm{T}}

\newcommand{\calJ}{\mathcal{J}}

\newcommand{\calN}{\mathcal{N}}

\newcommand{\bmNull}{{\mathbf{0}}}

\newcommand{\bmd}{{\mathbf{d}}}
\newcommand{\bme}{{\mathbf{e}}}

\newcommand{\bmg}{{\mathbf{g}}}

\newcommand{\bmq}{{\mathbf{q}}}

\newcommand{\bmx}{{\mathbf{x}}}
\newcommand{\bmy}{{\mathbf{y}}}
\newcommand{\bmz}{{\mathbf{z}}}

\newcommand{\bmC}{{\mathbf{C}}}
\newcommand{\bmD}{{\mathbf{D}}}

\newcommand{\bmG}{{\mathbf{G}}}
\newcommand{\bmH}{{\mathbf{H}}}
\newcommand{\bmI}{{\mathbf{I}}}

\newcommand{\bmY}{{\mathbf{Y}}}
\newcommand{\bmZ}{{\mathbf{Z}}}

\newcommand{\ol}{\overline}
\newcommand{\wt}{\widetilde}

\newcommand{\bmdelta}{\boldsymbol{\delta}}

\newcommand{\nmda}{{N_\text{mda}}}

\newcommand{\bmEN}{\mathbf{1}}

\newcommand{\bIenen}{\Big(\bmI-\frac{1}{N} \bmEN\bmEN^\rmT\Big)}

\begin{document}
\title{Accounting for model errors in iterative ensemble smoothers}
\author[12]{Geir Evensen}
\affil[1]{International Research Institute of Stavanger, Bergen, Norway}
\affil[2]{Nansen Environmental and Remote Sensing Center, Bergen, Norway}
\maketitle
\begin{abstract}
In the strong-constraint formulation of the history-matching problem, we assume that all the model errors relate to a selection of uncertain model input parameters.
One does not account for additional model errors that could result from, e.g., excluded uncertain parameters, neglected physics in the model formulation, the use of an approximate model
forcing, or discretization errors resulting from numerical approximations.
If parameters with significant uncertainties are unaccounted for, there is a risk for an unphysical update, of some uncertain parameters, that compensates for errors in the omitted parameters.
This paper gives the theoretical foundation for introducing model errors in ensemble methods for history matching.
In particular, we explain procedures for practically including model errors in iterative ensemble smoothers like ESMDA and IES, and we demonstrate the impact of adding
(or neglecting) model errors in the parameter-estimation problem. Also, we present a new result regarding the consistency of using the sample covariance of the predicted nonlinear measurements in the update schemes.
\end{abstract}

\section{Introduction\label{sec:intro}}

It is standard to assume the model to be perfect when we use ensemble smoothers for solving inverse problems.
This paper addresses the problem of consistently including the additional effect of stochastic model errors in different ensemble smoothers.
In particular, we consider methods such as Ensemble Smoother (ES) \citep[][]{eve96a,eve09book},
Iterative ES (IES) \citep{che12b,che13a},
and Ensemble Smoother with Multiple Data Assimilations (ESMDA) \citep{eme13a}. 

There is a vast literature on solving the data-assimilation problem in the presence of model errors \citep[see, e.g., the reviews by][]{car16a,har17a}. We traditionally characterize the data-assimilation problem as being either a weak-constraint or a strong-constraint
problem, dependent on whether we include the dynamical model as a strong or a weak constraint in the cost function.
The classical book by \citet{ben92c} is entirely devoted to solving the weak-constraint inverse problem, and it illustrates that in the weak-constraint case we need to invert simultaneously for the state vector as
a function of both space and time.

In \citet{ekn97a} the variational formulation was solved for a weak-constraint parameter and state estimation problem using the representer method by \citet{ben92c}, and in \citet{eve03a} different methods
including ensemble methods were used to solve a weak-constraint state- and parameter-estimation problem.
A conclusion from these works is that, if model errors are present, we need to increase the dimension of the problem by either updating the model solution as a function of space and time or by estimating the
actual model errors and after that solve for the model solution. It is also interesting to realize that the simultaneous state-parameter estimation problem becomes nonlinear even if the model itself is linear (except for in a very
trivial case), see for instance \citet{eve03a}.
Also, when using Ensemble Kalman Filter (EnKF) by \citet{eve94a} or ES, it is relatively easy to include stochastic model errors as long as we update both the parameters and the state variables simultaneously \citep{eve03a,eve09a}.

At this point, we should mention that we do not distinguish between model errors and model bias. In fact, with time-correlated stochastic model errors, and if the correlation becomes perfect (equal to one), then
the model errors become equivalent to a constant bias. Fortunately, the procedure outlined in this paper can be used to estimate both model errors and model bias.

In a recent paper by \citet{sak18a}, the Iterative EnKF was reformulated to allow for additive model errors. However, for the history-matching problem, we need to account for more general representations of the
error term since the solution of the reservoir simulator depends nonlinearly on the errors in the rate data used to force the model.

We will start by defining the standard strong-constraint history-matching problem, and after that, we move on to the general weak-constraint formulation. We then formally derive the ES, ESMDA, and IES, in the
presence of model errors. The different smoother methods are used in a simple example to demonstrate the consistency of the formulation used and to illustrate the impact of model errors on the parameter-estimation problem.

\section{Standard history-matching problem\label{sec:HMproblem}}

The strong-constraint formulation given by \citet{eve18b} is attractive because it simplifies the inverse problem and efficient ensemble smoothers can be defined.
A first fundamental assumption is that we have a perfect deterministic forward model where the prediction $\bmy$ only depends on the input model parametrization $\bmx$,
\begin{equation}
\bmy = \bmg(\bmx).
\label{eq:model}
\end{equation}
In a reservoir history-matching problem, the model operator is the reservoir simulation model, which predicts the observed production of oil, water, and gas, from the reservoir.
Thus, given the true parametrization of $\bmx$, the true prediction of $\bmy$ is precisely determined by the model in Eq.~(\ref{eq:model}).
Also, we have measurements $\bmd$ of the true $\bmy$ given as
\begin{equation}
 \bmd = \bmy + \bme.
\label{eq:measurements}
\end{equation}
From evaluating the model operator $\bmg(\bmx)$, given a realization of the uncertain model parameters $\bmx \in \Re^n$, we uniquely determine
a realization of predicted measurements $\bmy \in \Re^m$ (corresponding to the real measurements $\bmd \in \Re^m$). Here $n$ is the number of parameters and $m$ the number of measurements.
We want to use the measurements to estimate the parameters $\bmx$, and the measurements are assumed to contain random errors $\bme \in \Re^m$.

In history matching, it is common to define a prior distribution for the uncertain parameters since we usually will have more degrees of freedom in the parameters, than we have independent information
in the measurements.  Bayes' theorem gives the joint posterior pdf for  $\bmx$ and $\bmy$ as
\begin{equation}
\begin{split}
f(\bmx,\bmy\,|\,\bmd) &\propto f(\bmx,\bmy) f(\bmd\,|\,\bmy)\\
                  &= f(\bmx) f(\bmy\,|\,\bmx) f(\bmd\,|\,\bmy).
\label{eq:jointbayes}
\end{split}
\end{equation}
In the case of no model errors, the transition density $f(\bmy \,|\, \bmx)$  becomes the Dirac delta function, and we can write
\begin{equation}
 f(\bmx,\bmy\,|\,\bmd)  \propto f(\bmx) \delta(\bmy-\bmg(\bmx)) f(\bmd\,|\,\bmy).
\label{eq:strongjointbayes}
\end{equation}
We are interested in the marginal pdf for $\bmx$, which we obtain by integrating  Eq.~(\ref{eq:strongjointbayes}) over $\bmy$, giving
\begin{equation}
\begin{split}
 f(\bmx\,|\,\bmd) &\propto \int   f(\bmx) \bmdelta(\bmy-\bmg(\bmx)) f(\bmd\,|\,\bmy) \rmd\bmy \\
              &= f(\bmx) f(\bmd\,|\,\bmg(\bmx)) .
\end{split}
\label{eq:margbayes}
\end{equation}

When introducing the normal priors 
\begin{align}
f(\bmx)                       &= \calN(\bmx^\rmf, \bmC_{xx}) ,\label{eq:vecx}\\ 
f(\bmd \,|\, \bmg(\bmx))= f(\bme)       &= \calN(\bmNull, \bmC_{dd}) ,\label{eq:veceps}
\end{align}
we can write  Eq.~(\ref{eq:margbayes}) as
\begin{equation}
\begin{split}
f(\bmx\,|\,\bmd)&\propto  \exp  \bigg\{  -\ento \Big(\bmx - \bmx^\rmf\Big)^\rmT \bmC^{-1}_{xx}  \Big(\bmx - \bmx^\rmf\Big) \bigg\}  \\
            &\times   \exp  \bigg\{  -\ento \Big(\bmg(\bmx)-\bmd \Big)^\rmT \bmC^{-1}_{dd}  \Big(\bmg(\bmx)-\bmd \Big) \bigg\} .
\end{split}
\label{eq:marggauss}
\end{equation}
Note that the posterior pdf in Eq.~(\ref{eq:marggauss}) is non-Gaussian due to the nonlinear model $\bmg(\bmx)$.
Maximizing $f(\bmx\,|\,\bmd)$ is equivalent to minimizing the cost function
\begin{equation}
\begin{split}
 \calJ (\bmx) &=  \big(\bmx-\bmx^\rmf\big)^\rmT \bmC_{xx}^{-1} \big(\bmx-\bmx^\rmf\big) \\ &+
\big(\bmg(\bmx)-\bmd\big)^\rmT\bmC_{dd}^{-1}\big(\bmg(\bmx)-\bmd\big).
\label{eq:costfunction}
\end{split}
\end{equation}
Most methods for history matching apply the assumptions of a perfect model and Gaussian priors, and they attempt to solve
either one of Eqs.~(\ref{eq:marggauss}) or (\ref{eq:costfunction}).

For this strong-constraint problem, \citet{eve18b} explained how Eqs.~(\ref{eq:marggauss}) or (\ref{eq:costfunction}) can be
approximately solved using the ES, ESMDA, and IES.
We can interpret these methods to approximately sample the posterior pdf in Eq.~(\ref{eq:marggauss}), and we can easily derive them as an ensemble of minimizing solutions of the cost function in Eq.~(\ref{eq:costfunction})
written for each realization as
\begin{equation}
\begin{split}
 \calJ (\bmx_j) &=  \big(\bmx_j-\bmx_j^\rmf\big)^\rmT \bmC_{xx}^{-1} \big(\bmx_j-\bmx_j^\rmf\big) \\ & + \big(\bmg(\bmx_j)-\bmd_j\big)^\rmT\bmC_{dd}^{-1}\big(\bmg(\bmx_j)-\bmd_j\big).
\label{eq:costfja}
\end{split}
\end{equation}
This approach relates to the papers on Ensemble Randomized Likelihood (EnRML) \citep{kit95a,oli96a}.
Note that the minimizing solutions will not precisely sample the posterior non-Gaussian distribution. 

\section{ES solution}
We obtain the ES solution by first sampling the prior parameters $\bmx_j^\rmf$ and the perturbed measurements $\bmd_j$ from
\begin{align}
\bmx_j^\rmf & \sim \calN(\bmx^\rmf, \bmC_{xx}) ,\label{eq:pf}\\ 
\bmd_j      & \sim \calN(\bmd, \bmC_{dd}) .\label{eq:ef} 
\end{align}
We then compute the model predictions $\bmy_j^\rmf$ from 
\begin{equation}
\bmy_j^\rmf = \bmg(\bmx_j^\rmf). 
\label{eq:modelpf}   
\end{equation}
We define the covariance matrix between two vectors $\bmx$ and $\bmy$ as the expectation
\begin{equation}
\bmC_{xy}= \E{\big(\bmx-\E{\bmx}\big)\big(\bmy-\E{\bmy}\big)^\rmT} ,
\label{eq:covariance}
\end{equation}
while in the ensemble methods we introduce the sample mean
\begin{equation}
\ol{\bmx} = \frac{1}{N} \sum_{j=1}^N  \bmx_j,
\label{eq:smean}
\end{equation}
and the sample covariance between two arbitrary vectors $\bmx$ and $\bmy$ as
\begin{equation}
\ol{\bmC}_{xy}= \frac{1}{N-1}\sum_{j=1}^N \big(\bmx_j-\ol{\bmx}\big)\big(\bmy_j-\ol{\bmy}\big)^\rmT  .
\label{eq:scovariance}
\end{equation}
\citet{eve18b} derived the ES update equations for the parameters $\bmx_j^\rma$ and the predictions $\bmy_j^\rma$ as
\begin{align}
\bmx_j^\rma &= \bmx_j^\rmf + \ol{\bmC}_{xy} \lp  \ol{\bmC}_{yy} + \bmC_{dd} \rp^{-1} \Big(\bmd_j - \bmy_j^\rmf\Big)  , \label{eq:nxajupdate}  \\    
\bmy_j^\rma &= \bmg(\bmx_j^\rma),  \label{eq:modelpa}
\end{align}
but see also the alternative derivation in Secs.~\ref{sec:ES}--\ref{sec:ESalg} below regarding the consistency of using the sample covariance $\ol{\bmC}_{yy}$ in Eq.~(\ref{eq:nxajupdate}).
To compute the ES solution, we start from an initial ensemble of parameters $\bmx_j^\rmf$ and perturbed measurements $\bmd_j$.  First, we generate a prior ensemble prediction $\bmy_j^\rmf$ 
by evaluating the model in Eq.~(\ref{eq:modelpf}) for each realization $\bmx_j^\rmf$.  Then we compute the prior ensemble covariance between the parameters and the predicted measurements 
$\ol{\bmC}_{xy} \in \Re^{n\times m}$ and the ensemble covariance of the predicted measurements $\ol{\bmC}_{yy} \in \Re^{m\times m}$ and use them
in the ES update in Eq.~(\ref{eq:nxajupdate}). Finally, we can recompute the model prediction using the updated parameters $\bmx_j^\rma$ in Eq.~(\ref{eq:modelpa}).

As an alternative to solving the model in Eq.~(\ref{eq:modelpa}) for $\bmy_j^\rma$, it is possible to compute the updated prediction directly from an update equation
\begin{equation}
\bmy_j^\rma = \bmy_j^\rmf + \ol{\bmC}_{yy} \lp  \ol{\bmC}_{yy} + \bmC_{dd} \rp^{-1} \Big(\bmd_j - \bmy_j^\rmf\Big)  , 
\label{eq:yupdate} 
\end{equation}
and in the case with a linear model, the result would be identical to solving Eq.~(\ref{eq:modelpa}). However, due to the nonlinearity of the deterministic model, the Eqs.~(\ref{eq:modelpa}) and
(\ref{eq:yupdate}) will give different results for $\bmy_j^\rma $.  The Eq.~(\ref{eq:yupdate}) is just the standard ES update $\bmy_j^\rma$ given the prior forecast ensemble for $\bmy^\rmf_j$
and the perturbed measurements $\bmd_j$ of $\bmy$.  We will later show that, for nonlinear models, there may be a benefit of computing $\bmy_j^\rma$ indirectly through integration of the model in
Eq.~(\ref{eq:modelpa}), initialized with $\bmx_j^\rma$, rather than using the direct update in Eq.~(\ref{eq:yupdate}).  Also, the indirect update in Eq.~(\ref{eq:modelpa}) allows for the use of 
iterations.

\section{General weak constraint problem\label{sec:general}}
Lets now look at a formulation where we assume that the model depends nonlinearly on the model errors $\bmq$ as well as the parameters $\bmx$, i.e.,
\begin{equation}
\bmy = \bmg(\bmx,\bmq) .      
\label{eq:modies}
\end{equation}
The model operator can be somewhat general, including a recursion or time steps, and the model error is a vector of noise components that could represent, e.g., the time-corre\-lated noise in rate data used
to force a reservoir simulation model over a specific period. Thus, there is a significant difference between $\bmx$ and $\bmq$.  Firstly, $\bmx$ is a static parameter that does not change with time. Thus, once the
parameters $\bmx$ are estimated, we can use them in a future prediction simulation.  The model errors $\bmq$, on the other hand, will vary in time and are only estimated for the period of the inverse calculation. Only
in the case of time-correlated errors can the estimated $\bmq$ be used to some extent as a forcing  in the future prediction.

We assume that we have prior pdfs for the model errors and the parameters $f(\bmx)$ and $f(\bmq)$. It is common, but not necessary, to assume that the model errors and parameters are independent
so we can write the joint prior pdf as $f(\bmx,\bmq)=f(\bmx) f(\bmq)$.
The transition density for the model evolution is 
\begin{equation}
f(\bmy|\bmx,\bmq) = \delta  \big(\bmy-\bmg(\bmx,\bmq) \big),
\label{eq:modev}
\end{equation}
and we obtain the joint pdf by multiplying the prior with the transition density,
\begin{equation}
f(\bmy,\bmx,\bmq) = \delta  \big(\bmy-\bmg(\bmx,\bmq) \big) f(\bmx) f(\bmq) .
\label{eq:jointmodev}
\end{equation}
The likelihood function for the measurements given the prediction $\bmy$ is $f(\bmd|\bmy)$,
thus, the posterior conditional pdf becomes
\begin{equation}
 f(\bmy,\bmx,\bmq|\bmd) \propto f(\bmd|\bmy) \delta  \big(\bmy-\bmg(\bmx,\bmq) \big) f(\bmx) f(\bmq).
\label{eq:postiesa}
\end{equation}
Since $\bmy$ is given by the model in Eq.~(\ref{eq:modies}) as soon as we know $\bmx$ and $\bmq$, we can compute the marginal density 
\begin{equation}
\begin{split}
 f(\bmx,\bmq|\bmd) &\propto \int f(\bmd|\bmy) \delta  \big(\bmy-\bmg(\bmx,\bmq) \big) f(\bmx) f(\bmq) \rmd\bmy \\
                   &=      f\big(\bmd|\bmg(\bmx,\bmq)\big) f(\bmx) f(\bmq) .
\end{split}
\label{eq:postiesb}
\end{equation}

\subsection{Gaussian priors}
We now assume Gaussian priors
\begin{equation}
 f(\bmx) \propto \exp  \bigg\{ -\ento  \big(\bmx-\bmx^\rmf\big)^\rmT \bmC_{xx}^{-1} \big(\bmx-\bmx^\rmf\big)  \bigg\},
\label{eq:priorp}
\end{equation}
and
\begin{equation}
 f(\bmq) \propto \exp  \bigg\{ -\ento \big(\bmq-\bmq^\rmf\big)^\rmT \bmC_{qq}^{-1} \big(\bmq-\bmq^\rmf\big)   \bigg\},
\label{eq:priorq}
\end{equation}
where $\bmq^\rmf$ would normally be zero.
The likelihood for the measurements becomes
\begin{equation}
\begin{split}
f\big(\bmd&|\bmg(\bmx,\bmq)\big) = \\
    & \exp  \bigg\{ -\ento   \big(\bmg(\bmx,\bmq) -\bmd \big)^\rmT \bmC_{dd}^{-1} \big(\bmg(\bmx,\bmq) -\bmd\big) \bigg\} , 
\end{split}
\label{eq:likelihoodies}
\end{equation}
and we write the marginal posterior as 
\begin{equation}
\begin{split}
f(\bmx,\bmq|\bmd) \propto& \\
            \exp  \bigg\{&-\ento  \big(\bmx-\bmx^\rmf\big)^\rmT \bmC_{xx}^{-1} \big(\bmx-\bmx^\rmf\big)            \\
           &              -\ento  \big(\bmq-\bmq^\rmf\big)^\rmT \bmC_{qq}^{-1} \big(\bmq-\bmq^\rmf\big)            \\
           &              -\ento  \big(\bmg(\bmx,\bmq) -\bmd \big)^\rmT \bmC_{dd}^{-1} \big(\bmg(\bmx,\bmq) -\bmd\big) \bigg\} .
\end{split}
\label{eq:postiesc}
\end{equation}
Maximizing the posterior pdf in Eq.~(\ref{eq:postiesc}) is equivalent to minimizing the cost function
\begin{equation}
\begin{split}
 \calJ (\bmx,\bmq)&=  \big(\bmx-\bmx^\rmf\big)^\rmT \bmC_{xx}^{-1} \big(\bmx-\bmx^\rmf\big) \\
                  &+  \big(\bmq-\bmq^\rmf\big)^\rmT \bmC_{qq}^{-1} \big(\bmq-\bmq^\rmf\big) \\
& + \big(\bmg(\bmx,\bmq)-\bmd\big)^\rmT\bmC_{dd}^{-1}\big(\bmg(\bmx,\bmq)-\bmd\big).
\end{split}
\label{eq:costies}
\end{equation}
As in the strong constraint case we sample the priors and define a cost function for each sample realization, and we obtain the weak constraint analog to the strong-constraint cost function in Eq.~(\ref{eq:costfja}), i.e.,
\begin{equation}
\begin{split}
 \calJ (\bmx_j,\bmq_j)&=  \big(\bmx_j-\bmx_j^\rmf\big)^\rmT \bmC_{xx}^{-1} \big(\bmx_j-\bmx_j^\rmf\big) \\
                  &+   \big(\bmq_j-\bmq_j^\rmf\big)^\rmT \bmC_{qq}^{-1} \big( \bmq_j-\bmq_j^\rmf\big) \\
& + \big(\bmg(\bmx_j,\bmq_j)-\bmd_j\big)^\rmT\bmC_{dd}^{-1}\big(\bmg(\bmx_j,\bmq_j)-\bmd_j\big).
\end{split}
\label{eq:costj}
\end{equation}

\subsection{Stationarity condition}
To develop a consistent set of equations for the different methods, it is simpler to rewrite the cost function
for an augmented variable $\bmz^\rmT=(\bmx^\rmT,\bmq^\rmT)$. We can then define the covariance
\begin{equation}
 \bmC_{zz} = 
\begin{pmatrix}
\bmC_{xx} & \bmC_{xq} \\
\bmC_{qx} & \bmC_{qq} 
\end{pmatrix} ,
\label{eq:}
\end{equation}
where we allow for correlations between $\bmx$ and $\bmq$ since this correlation becomes important in the iterative methods,
and we can rewrite the cost function in Eq.~(\ref{eq:costj}) as
\begin{equation}
\begin{split}
 \calJ (\bmz_j)&=  \big(\bmz_j-\bmz_j^\rmf\big)^\rmT \bmC_{zz}^{-1} \big(\bmz_j-\bmz_j^\rmf\big) \\
& + \big(\bmg(\bmz_j)-\bmd_j\big)^\rmT\bmC_{dd}^{-1}\big(\bmg(\bmz_j)-\bmd_j\big).
\end{split}
\label{eq:costjz}
\end{equation}
This cost function is slightly more general than the one in Eq.~(\ref{eq:costj}) since we do not require independence betweeen $\bmq$ and $\bmx$.
By taking the gradient with respect to $\bmz_j$ we obtain
%
%
%
the following closed system of nonlinear equations for $\bmz_j$
\begin{align}
 \bmC_{zz}^{-1} (\bmz_j - \bmz_j^\rmf) & +  \nabla_{\bmz} \bmg  (\bmz_j) \bmC_{dd}^{-1} \big(\bmg(\bmz_j) - \bmd_j \big) = 0. \label{eq:ELza}
\end{align}
In the case of a linear model, the solution is the
standard Kalman filter update equation for the mean. However, the presence of the nonlinear function $\bmg(\bmz_j)$ makes the solution more elaborate.
All of ES, ESMDA and IES are developed to solve this system of equations.

\subsection{Derivation of ES update equations\label{sec:ES}}
To derive the ES solution, we will use a Taylor expansion and approximation that allows us to obtain a
closed form solution for each realization of $\bmz_j$, i.e.,
\begin{align}
 \bmg(\bmz_j) &= \bmg(\bmz_j^\rmf) + \gradgzfj (\bmz_j - \bmz_j^\rmf),
\label{eq:taylorBz}
\end{align}
where we have defined
\begin{align}
 \gradgzfj^\rmT & = \nabla_\bmz \bmg  (\bmz)\big|_{\bmz=\bmz_j^\rmf} .
\label{eq:modgrad}
\end{align}
Thus, we approximate the nonlinear function $\bmg(\bmz_j)$ with its linearization in Eq.~(\ref{eq:taylorBz}) around $\bmz=\bmz_j^\rmf$ and in addition evaluate the gradient in Eq.~(\ref{eq:ELza}) at $\bmz_j^\rmf$.
We now have the gradient of the model, $\gradgzfj$ defined in Eq.~(\ref{eq:modgrad}), which differ for each realization. We wish to replace the individual model gradients with an ``averaged'' gradient that is common 
for all realizations, and for now, we denote it $\bmG$, which allows us to write Eq.~(\ref{eq:ELza}) as
\begin{align}
 \big( \bmC_{zz}^{-1}  + \bmG^\rmT  \bmC_{dd}^{-1} \bmG \big) \big(\bmz_j - \bmz_j^\rmf\big) 
                                   & = \bmG^\rmT  \bmC_{dd}^{-1} \big(\bmd_j - \bmg(\bmz_j^\rmf)\big) . \label{eq:ELzb}
\end{align}
By solving for $\bmz_j$ and using the matrix identity
\begin{equation}
 \big(\bmG^\rmT \bmD^{-1} \bmG + \bmC^{-1} \big)^{-1} \bmG^\rmT \bmD^{-1} = \bmC \bmG^\rmT \big(\bmG \bmC\bmG^\rmT + \bmD \big)^{-1} ,
\label{eq:woodbury}
\end{equation}
(which can be derived from the Woodbury identity), where we substitute $\bmC_{zz}$ for $\bmC$, and $\bmC_{dd}$ for $\bmD$, we obtain the solution for $\bmz_j$ as
\begin{align}
 \bmz_j & = \bmz_j^\rmf +  \bmC_{zz} \bmG^\rmT \Big(\bmG \bmC_{zz}\bmG^\rmT + \bmC_{dd}    \Big)^{-1} \Big(\bmd_j - \bmg(\bmz_j^\rmf)  \Big) .
\label{eq:ESza} 
\end{align}

\subsection{Linear regression for $\bmG$}
\citet{eve18b} used a Taylor expansion of $\bmg(\bmz_j)$ around the ensemble mean $\ol{\bmz}=\bmz^\rmf$
and could then replace the individual gradients in Eq.~(\ref{eq:modgrad}) evaluated at $\bmz_j$ with the gradient $\bmG_{\ol{\bmz}}$ evaluated at the ensemble mean $\ol{\bmz}$.
He then showed that $\ol{\bmC}_{zy} \approx \bmG_{\ol{\bmz}} \ol{\bmC}_{zz}$ and $\ol{\bmC}_{yy} \approx \bmG_{\ol{\bmz}} \ol{\bmC}_{zz}\bmG_{\ol{\bmz}}^\rmT$  \citep[][Eqs.~29 and ~30]{eve18b}, and he replaced the 
analytical gradients with ensemble covariances.

We will here use an interpretation based on linear regression \citep[see also][]{rey06a,che13a} 
where we start by defining
\begin{equation}
 \bmG \triangleq \bmC_{yz} \bmC_{zz}^{-1} .
\label{eq:Gzdef}
\end{equation}
Thus, we view $\bmG$ as the sensitivity matrix in a linear regression between $\bmy$ and $\bmz$ as
\begin{equation}
\bmC_{yz} =  \bmG \bmC_{zz}.
\label{eq:linreg}
\end{equation}
When we introduce the ensemble-anomaly matrices
\begin{align}
 \bmY &=\frac{1}{\sqrt{N-1}}\big(\bmy_1 - \ol{\bmy}, \ldots , \bmy_N - \ol{\bmy} \big) \in \Re^{m\times N}, \label{eq:Ysamp} \\
 \bmZ &=\frac{1}{\sqrt{N-1}}\big(\bmz_1 - \ol{\bmz}, \ldots , \bmz_N - \ol{\bmz} \big) \in \Re^{n\times N}, \label{eq:Zsamp} 
\end{align}
we can write Eq.~(\ref{eq:linreg}) as
\begin{equation}
\bmY\bmZ^\rmT \approx  \bmG \bmZ\bmZ^\rmT,
\label{eq:linregsampcov}
\end{equation}
or equivalently
\begin{equation}
\bmY \approx \bmG \bmZ,
\label{eq:linregsamp}
\end{equation}
where the approximation is the use of a finite ensemble size.
From Eq.~(\ref{eq:Gzdef}) we have 
\begin{equation}
\begin{split}
 \bmG 
           &\approx  \ol{\bmG} =\ol{\bmC}_{yz} (\ol{\bmC}_{zz})^{+} = \bmY\bmZ^\rmT \big( \bmZ \bmZ^\rmT \big)^+ = \bmY \bmZ^+,
\end{split}
\label{eq:Gzdefpseudo}
\end{equation}
where the superscript $^+$ denotes pseudo inverse. The unbiasedness of
$\ol{\bmG}$ can be shown, i.e.,
\begin{equation}
 \E{\ol{\bmG}}=\bmG .
\label{eq:unbiased}
\end{equation}

Also, using Eq.~(\ref{eq:Gzdef}), we can write 
\begin{equation}
\bmG \bmC_{zz} \bmG^\rmT  = \bmC_{yz} \bmC_{zz}^{-1}  \bmC_{zy} ,
\label{eq:cyy}
\end{equation}
and when  we introduce the ensemble representation from Eq.~(\ref{eq:Gzdefpseudo}), we obtain
\begin{equation}
\begin{split}
\ol{\bmG}\; \ol{\bmC}_{zz} \ol{\bmG}^\rmT 
                                          & = \bmY (\bmZ^+ \bmZ) (\bmZ^+ \bmZ) \bmY^\rmT .
\end{split}
\label{eq:Gyzzzzy}
\end{equation}
The projection $\bmZ^+ \bmZ$ is just the orthogonal projection onto the range of $\bmZ^\rmT$.
We will now  consider three cases:

\subsubsection{Linear model and measurement operator}
For a linear model and measurement operator we can write, e.g., $\bmY=\bmH \bmZ$, and
Eq.~(\ref{eq:Gyzzzzy}) becomes
\begin{equation}
\begin{split}
 \ol{\bmG}\; \ol{\bmC}_{zz} \ol{\bmG}^\rmT  &= \bmH \bmZ (\bmZ^+ \bmZ) (\bmZ^+ \bmZ) \bmZ^\rmT \bmH^\rmT \\
                                          &= \bmH \bmZ  (\bmH \bmZ)^\rmT = \bmY \bmY^\rmT =\ol{\bmC}_{yy} .
\end{split}
\label{eq:Hyzzzzy}
\end{equation}
Hence, Eq.~(\ref{eq:Hyzzzzy}) is consistent with the definition of the covariance matrix $\ol{\bmC}_{yy}= \bmY \bmY^\rmT$ for linear models and all combinations of $n$ and $N$.

\subsubsection{Nonlinear model and $n\geq N-1$}
In the case with a nonlinear model and $n\geq N-1$ the rank of $\bmZ$ is $N-1$, and the projection $\bmZ^+ \bmZ = \bIenen $ with $\bmEN \in \Re^{N}$ being a vector with all elements equal to one
\citep[see the Appendix in][]{sak12a}. This result is seen from the fact that $\bmZ$ has only one singular value equal to zero corresponding to
the right singular vector $\bmEN/\sqrt{N}$.  The projection $\bmZ^+ \bmZ$ is then just the  subtraction of the ensemble mean. Since we have already removed the ensemble mean from $\bmY$, we can write Eq.~(\ref{eq:Gyzzzzy}) as
\begin{equation}
\begin{split}
 \ol{\bmG}\; \ol{\bmC}_{zz} \ol{\bmG}^\rmT  &= \bmY \bIenen \bIenen \bmY^\rmT \\
                                          &= \bmY \bmY^\rmT = \ol{\bmC}_{yy} ,
\end{split}
\label{eq:Gyy}
\end{equation}
and as in the linear case, Eq.~(\ref{eq:Gyzzzzy}) exactly corresponds to the definition of the ensemble covariance $\ol{\bmC}_{yy}$.
This nonlinear case with  $n \geq N-1$ is the most considered case for history matching, data assimilation with nonlinear measurement operators, and iterative smoothers
used for sequential data assimilation in nonlinear models. Thus, for most applications of ensemble methods, we can replace the product $\ol{\bmG}\; \ol{\bmC}_{zz} \ol{\bmG}^\rmT$ with the 
sample covariance $\ol{\bmC}_{yy}$.  

\subsubsection{Nonlinear model and $n< N-1$}
In the case of a nonlinear model and $n < N-1$, which applies for the example considered in Section~\ref{sec:ex}, the expression in Eq.~(\ref{eq:Gyzzzzy}) is not equal to $\bmY\bmY^\rmT$ and we 
must include the projection and redefine the sample covariance as
\begin{equation}
\begin{split}
 \wt{\bmC}_{yy} 
                & \triangleq  \bmY (\bmZ^+ \bmZ) \big(   \bmY (\bmZ^+ \bmZ) \big)^\rmT,
\end{split}
\label{eq:defCyy}
\end{equation}
i.e., we compute the covariance of the predicted measurement anomalies projected onto the range of $\bmZ^\rmT$.
This case also applies for \textit{nonlinear models in the limit of infinite ensemble size.}
Thus, we must use the definition Eq.~(\ref{eq:defCyy}) to evaluate $\wt{\bmC}_{yy}$
to ensure consistency in the derivation of the update equation.

\subsection{ES algorithm\label{sec:ESalg}}
We can replace the gradient $\gradgzf$  in the update Eq.~(\ref{eq:ESza}), using Eqs.~(\ref{eq:linreg}) and (\ref{eq:cyy})
to obtain
\begin{align}
 \bmz_j^\rma & = \bmz_j^\rmf +  \bmC_{zy} \Big( \bmC_{yz}\bmC_{zz}^{-1} \bmC_{zy} + \bmC_{dd}    \Big)^{-1} \Big(\bmd_j - \bmg(\bmz_j^\rmf)  \Big).
\label{eq:ESzz} 
\end{align}
The solution of this equation is identical to the solution of Eq.~(\ref{eq:ELzb}) with $\bmG$ defined by Eq.~(\ref{eq:Gzdef}). However, if we replace 
$\bmC_{yz}\bmC_{zz}^{-1} \bmC_{zy}$ with $\bmC_{yy}$ in Eq.~(\ref{eq:ESzz}), the solutions of Eqs.~(\ref{eq:ELzb}) and (\ref{eq:ESzz}) will differ in the nonlinear case. 

By representing the covariances using a finite sample size and sample covariance matrices, we
get the update equation for the finite ensemble of $N$ realizations as
\begin{align}
 \bmz_j^\rma & = \bmz_j^\rmf +  \ol{\bmC}_{zy} \Big( \wt{\bmC}_{yy} + \bmC_{dd}    \Big)^{-1} \Big(\bmd_j - \bmg(\bmz_j^\rmf)  \Big),
\label{eq:ESz} 
\end{align}
where  $\wt{\bmC}_{yy}$ is defined in (\ref{eq:defCyy}).
If we omit the projection in Eq.~(\ref{eq:defCyy}), then the solution computed by Eqs.~(\ref{eq:ESzz}) (using sample covariances)
and (\ref{eq:ESz}) will differ in the case of a nonlinear model and $n<N-1$, also when $N$ becomes infinitely large. Thus, our update will be biased.

To compute the ES update, we start by sampling the Gaussian prior variables for the parameters $\bmx^\rmf_j$, the model errors $\bmq_j^\rmf$, and the measurement
perturbations $\bme_j$,
\begin{align}
\bmx_j^\rmf           & \sim \calN(\bmx^\rmf, \bmC_{xx}^\rmf)                        ,\label{eq:pdfp}\\ 
\bmq_j^\rmf           & \sim \calN(\bmNull, \bmC_{qq}^\rmf)                          ,\label{eq:pdfq}\\ 
\bme_j           & \sim \calN(\bmNull, \bmC_{dd})                          .\label{eq:pdfd}
\end{align}
The vector $\bmq$ contains all stochastic model errors over the time interval of the model integration, and the errors can also have correlations in time.
We obtain the model prediction from the model written on the form
\begin{equation}
\bmy_j^\rmf = \bmg(\bmx_j^\rmf,\bmq_j^\rmf) , 
\label{eq:generalmod}
\end{equation}
where the model operator depends nonlinearly on the model-error term. 
Next, we can compute the sample covariances   $\ol{\bmC}_{xy}^\rmf$,  and $\ol{\bmC}_{qy}^\rmf$,  from the ensembles of $\bmy_j^\rmf$, $\bmx_j^\rmf$, and $\bmq_j^\rmf$, and 
$\wt{\bmC}_{yy}^\rmf$ from the definition in Eq.~(\ref{eq:defCyy}).

Eq.~(\ref{eq:ESz}) defines the final update equations for $\bmx_j^\rma$  and $\bmq_j^\rma$ which becomes 
\begin{align}
\bmx_j^\rma &= \bmx_j^\rmf + \ol{\bmC}_{xy}^{\rmf} \big(\wt{\bmC}_{yy}^{\rmf} + \bmC_{dd}\big)^{-1} \Big(\bmd_j - \bmy_j^\rmf\Big), \label{eq:pupdatege}   \\
\bmq_j^\rma &= \bmq_j^\rmf + \ol{\bmC}_{qy}^{\rmf} \big(\wt{\bmC}_{yy}^{\rmf} + \bmC_{dd}\big)^{-1} \Big(\bmd_j - \bmy_j^\rmf\Big). \label{eq:qupdatege}  
\end{align}
We then rerun the model using the updates $\bmx_j^\rma$  and $\bmq_j^\rma$ to get
\begin{align}
\bmy^\rma_{j} &= \bmg(\bmx^\rma_j,\bmq^\rma_{j}) .
 \label{eq:generalmoda}
\end{align}
Alternatively, we can also compute the update of the predicted measurements from 
\begin{equation}
\bmy_j^\rma = \bmy_j^\rmf + \ol{\bmC}_{yy}^{\rmf} \big(\ol{\bmC}_{yy}^{\rmf} + \bmC_{dd}\big)^{-1} \Big(\bmd_j - \bmy_j^\rmf\Big),  \label{eq:yupdatege}  
\end{equation}
and in the case of a linear model the result would be identical to that obtained by integrating the model in Eq.~(\ref{eq:generalmoda}).

\section{Iterative smoothers in the presence of model errors\label{sec:weakiterative}}
The critical approximations used in the derivation of ES are,
firstly, the linearization in Eq.~(\ref{eq:taylorBz}) of the model about $\bmx_j^\rmf$ meaning that large updates will have large errors,
and secondly, that an averaged statistical ensemble gradient replaces the exact analytic gradients. Only a single linear update step is computed, and with strong nonlinearities, these approximations may
lead to poor results as was discussed in \citet{eve18b}. 

The minimization problem in Eq.~(\ref{eq:costfja}) can be solved using iterative methods like IES, ESMDA, and IEnKF.
The iterative ensemble smoother (IES) by \citet{che12b,che13a} minimizes the ensemble of cost functions by direct minimization using an approximate ensemble gradient.
Alternatively, the Ensemble Smoother with Multiple Data Assimilations (ESMDA) by \citet{eme12a} re\-writes the ES update equation as a sequence of recursive updates by inflating the measurement errors, and thereby reduces
the approximation introduced by the linearization in the ES update.

\citet{sak12a,boc14a} derived the iterative EnKF (IEnKF) and iterative Ensemble Kalman Smoother (IEnKS) to better handle nonlinearities in the dynamical model and the observation operator. The focus was on state estimation 
where the model state at the time $t_i$ is updated using measurements of the state at time $t_{i+1}$.  IEnKF solves the same kind of problem as given by the marginal conditional
pdf in Eq.~(\ref{eq:margbayes}) or the cost function in Eq.~(\ref{eq:costfunction}).
In a recent paper by \citet{sak18a} the IEnKF was extended to account for additive model errors, and the method should also be applicable for the history matching problem in the presence
of additive model errors.

In the following, we will present variants of ESMDA and IES that take more general
model errors into account as is required when solving the weak constraint history-matching problem.

\subsection{ESMDA\label{sec:esmda}}
As explained in \citet{eve18b}, ESMDA solves the standard ES update equations using a predefined number of recursive steps. In each step, the measurement error covariance and associated
measurement perturbations are inflated to reduce the impact of the measurements. With correctly chosen inflation factors and a linear model and observation operators,
the ESMDA update precisely replicates the ES update.
When the model or observation operators are nonlinear, it turns out that the use of multiple short update steps reduces the errors and improves the solution as
compared to using one long update step in ES. 

From the previous discussion, it is clear that, in the presence of model errors, we need to recursively update both the parameters and the model errors.
It is easiest to derive ESMDA by using a tempering of the likelihood function \citep{nea96a} which leads to a 
recursive minimization of a sequence of $N_\textrm{mda}$ cost functions, \citep{sto15a,eve18b},
\begin{equation}
\begin{split}
 \calJ (\bmz_{j,i+1})&=  \big(\bmz_{j,i+1}-\bmz_{j,i}\big)^\rmT \big(\bmC_{zz}^{i}\big)^{-1} \big(\bmz_{j,i+1}-\bmz_{j,i}\big) \\
& + \big(\bmg(\bmz_{j,i+1})-\bmd-\sqrt{\alpha_{i+1}}\bme_{j,i}\big)^\rmT 
  \big(\alpha_{i+1}\bmC_{dd}\big)^{-1}\big(\bmg(\bmz_{j,i+1})-\bmd -\sqrt{\alpha_{i+1}}\bme_{j,i}\big),
\end{split}
\label{eq:costjza}
\end{equation}
where we evaluate $\bmC^i_{zz}$ at the $i$th iterate $\bmz_i$, and we must have 
\begin{equation}
\sum_{i=1}^{\nmda} \frac{1}{\alpha_i} = 1 .
\label{eq:alphasumb}
\end{equation}
Similarly to the derivation of the ES in the previous section, we obtain the recursive update equations for ESMDA given by Eqs.~(\ref{eq:enmdax}) and (\ref{eq:esmdaq}) in the algorithm below.
As in ES, the update direction is computed based on a linearization around the prior realizations of each update step. Thus, we can interpret the ES update as
taking one long Euler step of length $\Delta \tau =1$ in pseudo time $\tau$, while in ESMDA we take a predefined number of shorter Euler steps of step length $\Delta \tau_i = 1/\alpha_i$ that satisfy
Eq.~(\ref{eq:alphasumb}) \citep[see e.g. the discussion in][]{eve18b}.

To compute the ESMDA solution, we start by sampling the initial ensembles from Eqs.~(\ref{eq:pdfp}) and (\ref{eq:pdfq}) to initialize the recursion in ESMDA 
\begin{align}
\bmx_{j,0}           & \sim \calN(\bmx^\rmf, \bmC_{xx})                        ,\label{eq:pdfpmda}\\ 
\bmq_{j,0}           & \sim \calN(\bmNull, \bmC_{qq})                          .\label{eq:pdfqmda}
\end{align}
Then the model is integrated according to Eq.~(\ref{eq:generalmod}) to obtain the prior ensemble prediction for the first ESMDA step,
\begin{align}
\bmy_{j,0} &= \bmg(\bmx_{j,0},\bmq_{j,0}) , 
\label{eq:generalmodamda}
\end{align}
and we compute recursively the following for each iteration $i=0 ,\ldots,N_\textrm{mda}-1$:

We construct the sample covariances   $\ol{\bmC}_{xy}^{\,i}$,  and $\ol{\bmC}_{qy}^{\,i}$,  from the ensembles of $\bmy_{j,i}$, $\bmx_{j,i}$, and $\bmq_{j,i}$, 
and $\wt{\bmC}_{yy}^i$ from the definition in Eq.~(\ref{eq:defCyy}),
and we sample the measurement perturbations
\begin{equation}
\bme_{j,i}        \sim \calN(\bmNull, \alpha_{i+1} \bmC_{dd}),
\label{eq:emda}
\end{equation}
used to generate the perturbed measurements 
\begin{equation}
 \bmd_{j,i} = \bmd + \bme_{j,i}
\label{eq:dpert}
\end{equation}
We then compute the updates 
\begin{align}
 \bmx_{j,i+1} &= \bmx_{j,i}  +  \ol{\bmC}^{\,i}_{xy} \Big( \wt{\bmC}^{\,i}_{yy} + \alpha_{i+1} \bmC_{dd}\Big)^{-1}  \Big( \bmd_{j,i} - \bmy_{j,i}\Big) , \label{eq:enmdax}\\ 
 \bmq_{j,i+1} &= \bmq_{j,i}  +  \ol{\bmC}^{\,i}_{qy} \Big( \wt{\bmC}^{\,i}_{yy} + \alpha_{i+1} \bmC_{dd}\Big)^{-1}  \Big( \bmd_{j,i} - \bmy_{j,i}\Big) , \label{eq:esmdaq}
\end{align}
and rerun the model to obtain the updated prediction
\begin{align}
\bmy_{j,i+1} &= \bmg(\bmx_{j,i+1},\bmq_{j,i+1}) , 
\label{eq:generalmodamdai}
\end{align}
for step $i+1$. We repeat this procedure until $i= N_\textrm{mda}-1$, which results in the ESMDA solution for $\bmx_j$, $\bmq_j$ and $\bmy_j$.


\subsection{IES\label{sec:ies}}
In IES we use a gradient based minimization method, and we need to evaluate the first and second order derivatives of the cost function in Eq.~(\ref{eq:costjz}) with respect to $\bmz$.
The gradient of the cost function in Eq.~(\ref{eq:costjz}) is already derived above as Eq.~(\ref{eq:ELza})
\begin{equation}
\nabla_\bmz \calJ(\bmz_j) =  \bmC_{zz}^{-1} (\bmz_j - \bmz_j^\rmf)  +  \nabla_\bmz \bmg  (\bmz_j) \bmC_{dd}^{-1} \big(\bmg(\bmz_j) - \bmd_j \big) .
\label{eq:gradIES}
\end{equation}
An approximation to the Hessian of the cost function is obtained by operating again by $\nabla_\bmz$ on the gradient in Eq.~(\ref{eq:gradIES}) to obtain
\begin{equation}
\nabla_\bmz \nabla_\bmz \calJ(\bmz_j) \approx  \bmC_{zz}^{-1}  +  \nabla_\bmz \bmg  (\bmz_j) \bmC_{dd}^{-1}  \big(\nabla_\bmz \bmg(\bmz_j)\big)^\rmT ,
\label{eq:hessIES}
\end{equation}
where we have neglected the second derivatives or Hessian of the vector function $\bmg(\bmz)$, i.e., $\nabla_\bmz \nabla_\bmz\bmg(\bmz)$.
We can then write a Gauss-Newton iteration 
\begin{equation}
 \bmz_{j,i+1} =\bmz_{j,i} - \gamma  \Delta\bmz_{j,i} ,
\label{eq:IESit}
\end{equation}
where we define $\Delta \bmz$ as the gradient normalized by the approximate Hessian as follows
\begin{align}
\Delta\bmz_{j,i}  &= \Big(\nabla_\bmz \nabla_\bmz \calJ(\bmz_{j,i})\Big)^{-1} \nabla_\bmz \calJ(\bmz_{j,i})  \nonumber\\
               &=\Big(  \bmC_{zz}^{-1}  +  \bmG_{j,i}^\rmT \bmC_{dd}^{-1}  \bmG_{j,i} \Big)^{-1} \bmC_{zz}^{-1} (\bmz_{j,i} - \bmz_j^\rmf) \label{eq:Deltaz}\\
               &+ \Big(  \bmC_{zz}^{-1}  +  \bmG_{j,i}^\rmT \bmC_{dd}^{-1}  \bmG_{j,i} \Big)^{-1} \bmG_{j,i}^\rmT \bmC_{dd}^{-1} \big(\bmg(\bmz_{j,i}) - \bmd_j \big),\nonumber
\end{align}
and we define
\begin{equation}
\bmG^\rmT_{i,j} = \nabla_\bmz \bmg  (\bmz_{i,j})
\label{eq:defGij}
\end{equation}
as the gradient of the model, evaluated at iteration $i$ and for ensemble member $j$.
The Eqs.~(\ref{eq:IESit}) and (\ref{eq:Deltaz}) defines the Ensemble Randomized Likelihood method \citep{kit95a,oli96a}.

Since we are not computing the analytical gradient of the model we will need to 
aproximate the ensemble of gradients with an averaged gradient like $\gradgzf$ from Eq.~(\ref{eq:Gzdefpseudo}), or we can evaluate the gradient at the ensemble average for the local iterate
$\bmG_{\ol{\bmz}}$ as was explained by \citet{eve18b}. The same model gradient is now used for all realizations, and this leads to a different solution than
the solution of the originally posed problem.

Eq.~(\ref{eq:Deltaz}) is exactly Eq.~(2) in \citet{che13a}. Also, \citet{che13a} suggested using the state covariance in the Hessian evaluated at the local iterate to simplify further
computations, since changing the Hessian does not change the gradient and thus the final converged solution (although it changes the step lengths in each iteration).

Thus, we can rewrite Eq.~(\ref{eq:Deltaz}) with the averaged model gradient and introduce the state covariance for the local iterate in the Hessian, to obtain
\begin{equation}
\begin{split}
\Delta\bmz_{j,i}
              &=\Big(  \big(\bmC^i_{zz}\big)^{-1}  +  \bmG_{i}^\rmT \bmC_{dd}^{-1}  \bmG_i \Big)^{-1} \bmC_{zz}^{-1} (\bmz_{j,i} - \bmz_j^\rmf) \\
              &+ \Big( \big(\bmC^i_{zz}\big)^{-1}  +  \bmG_{i}^\rmT \bmC_{dd}^{-1}  \bmG_{i} \Big)^{-1} \bmG_i^\rmT \bmC_{dd}^{-1} \big(\bmg(\bmz_{j,i}) - \bmd_j \big). 
\end{split}
\label{eq:DeltazB}
\end{equation}
Then using the corollaries 
\begin{align}
&  \lp \bmC^{-1} + \bmG^\rmT \bmD^{-1} \bmG\rp^{-1} = \bmC - \bmC \bmG^\rmT ( \bmG\bmC \bmG^\rmT + \bmD)^{-1} \bmG\bmC, \\
& \big(\bmG^\rmT \bmD^{-1} \bmG + \bmC^{-1} \big)^{-1} \bmG^\rmT \bmD^{-1} = \bmC \bmG^\rmT \big(\bmG \bmC\bmG^\rmT + \bmD \big)^{-1} ,
\label{eq:lemmas}
\end{align}
which are derived from the Woodbury identity, we can write Eq.~(\ref{eq:DeltazB}) as
\begin{equation}
\begin{split}
\Delta\bmz_{j,i} & = \bmC^i_{zz} \bmC_{zz}^{-1}(\bmz_{j,i} - \bmz^\rmf_j) \\
                 & - \bmC^i_{zz} \bmG_i^\rmT \Big(  \bmG_{i} \bmC^i_{zz} \bmG_i^\rmT +  \bmC_{dd} \Big)^{-1}  
                  \Big( \bmG_{i} \bmC^i_{zz} \bmC_{zz}^{-1} (\bmz_{j,i} - \bmz_j^\rmf) - \big( \bmg(\bmz_{j,i})-\bmd_j \big) \Big).
\end{split}
\label{eq:DeltazB1}
\end{equation}
Now, from Eqs.~(\ref{eq:linreg}) and (\ref{eq:cyy}) we can write Eq.~(\ref{eq:DeltazB1}) as
\begin{equation}
\begin{split}
\Delta\bmz_{j,i} & = \bmC^i_{zz} \bmC_{zz}^{-1}(\bmz_{j,i} - \bmz^\rmf_j) \\
                 & - \bmC^i_{zy} \Big(  \bmC^i_{yz} (\bmC_{zz}^i)^{-1}  \bmC^i_{zy} +  \bmC_{dd} \Big)^{-1}  
                  \Big( \bmC^i_{yz} \bmC_{zz}^{-1} (\bmz_{j,i} - \bmz_j^\rmf) - \big( \bmg(\bmz_{j,i})-\bmd_j \big) \Big).
\end{split}
\label{eq:DeltazB2}
\end{equation}
In the original algorithm the expression $\bmC^i_{yz} (\bmC_{zz}^i)^{-1}  \bmC^i_{zy}$ was replaced with the covariance $\bmC^i_{yy}$. But, as we have seen, this
will break the consistency between Eqs.~(\ref{eq:DeltazB}) and (\ref{eq:DeltazB2}) in the nonlinear case with $n<N-1$, and we need to use the definition in Eq.~(\ref{eq:defCyy}) to replace this expression.

The numerical solution method for this equation is discussed in more detail by \citet{che13a}. It is clear that it is the introduction of low-rank ensemble representations of the covariances that makes
it possible to compute the update steps $\Delta\bmz_{j,i}$, and the computation requires the use of singular-value decompositions and pseudo inversions.

\begin{figure*}[p]
\begin{center}
\tabcolsep=0.0pt
\begin{tabular}{cc}
\includegraphics[width=0.42\textwidth,trim= 0 0 0 0]{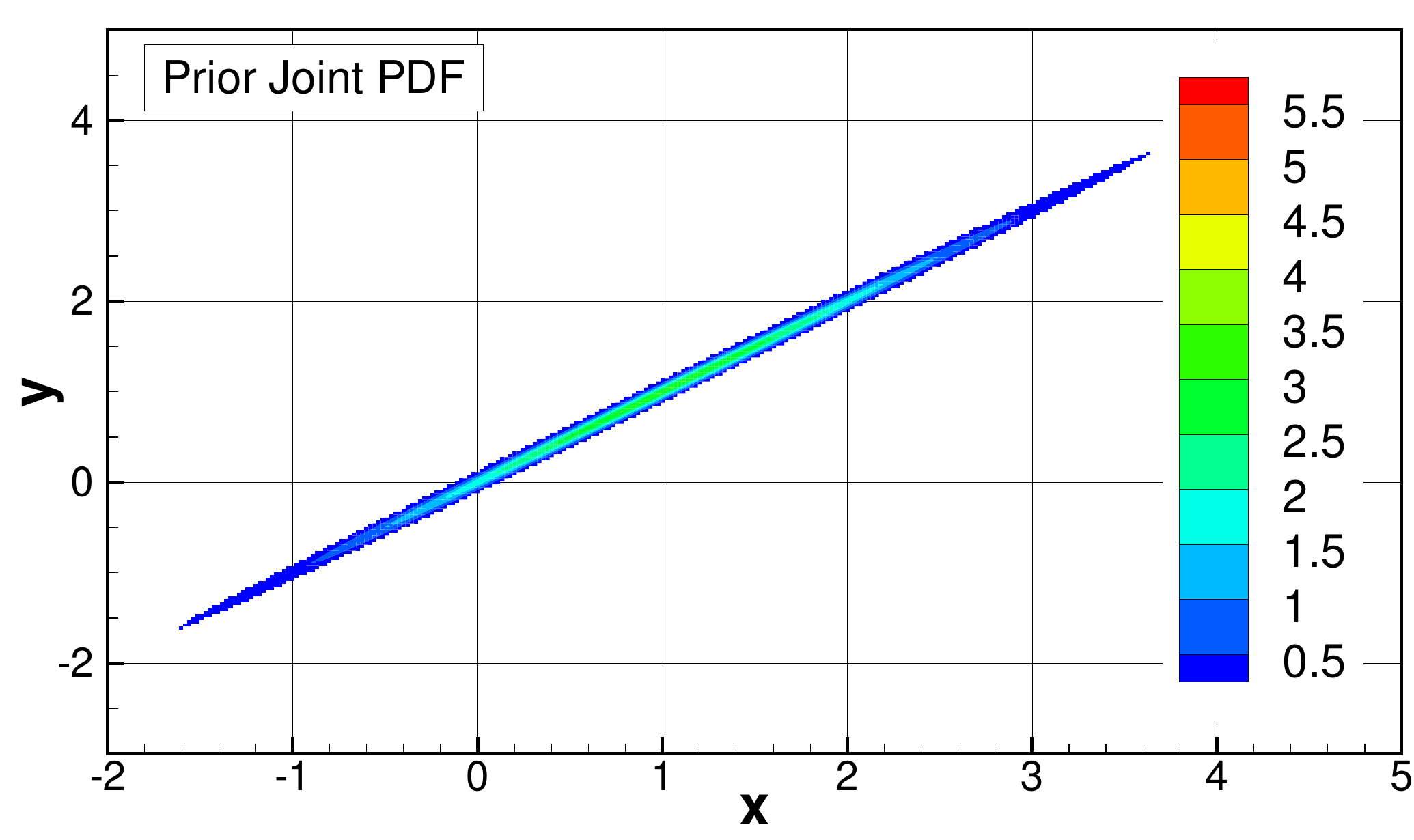}&
\includegraphics[width=0.42\textwidth,trim= 0 0 0 0]{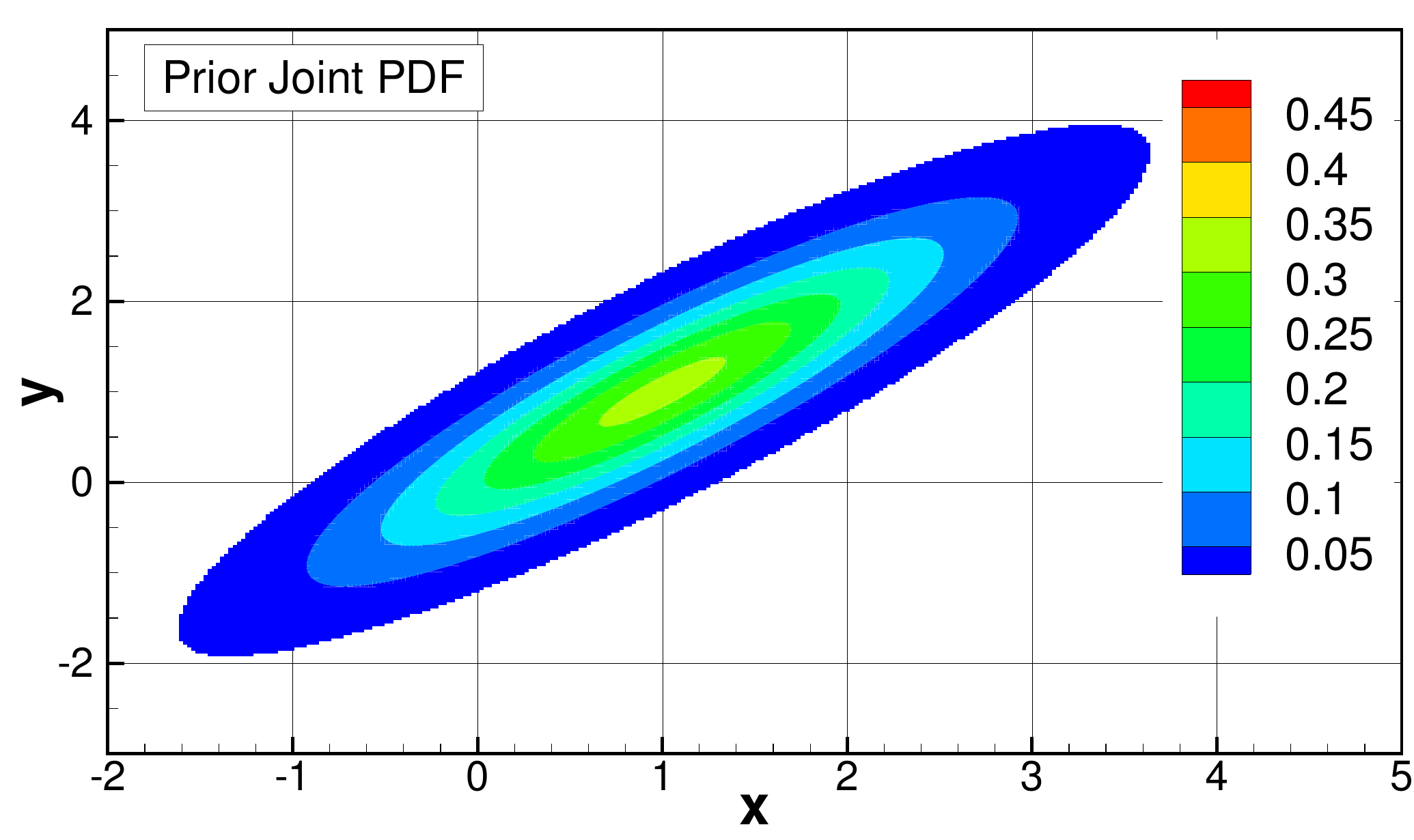}\\
\includegraphics[width=0.42\textwidth,trim= 0 0 0 0]{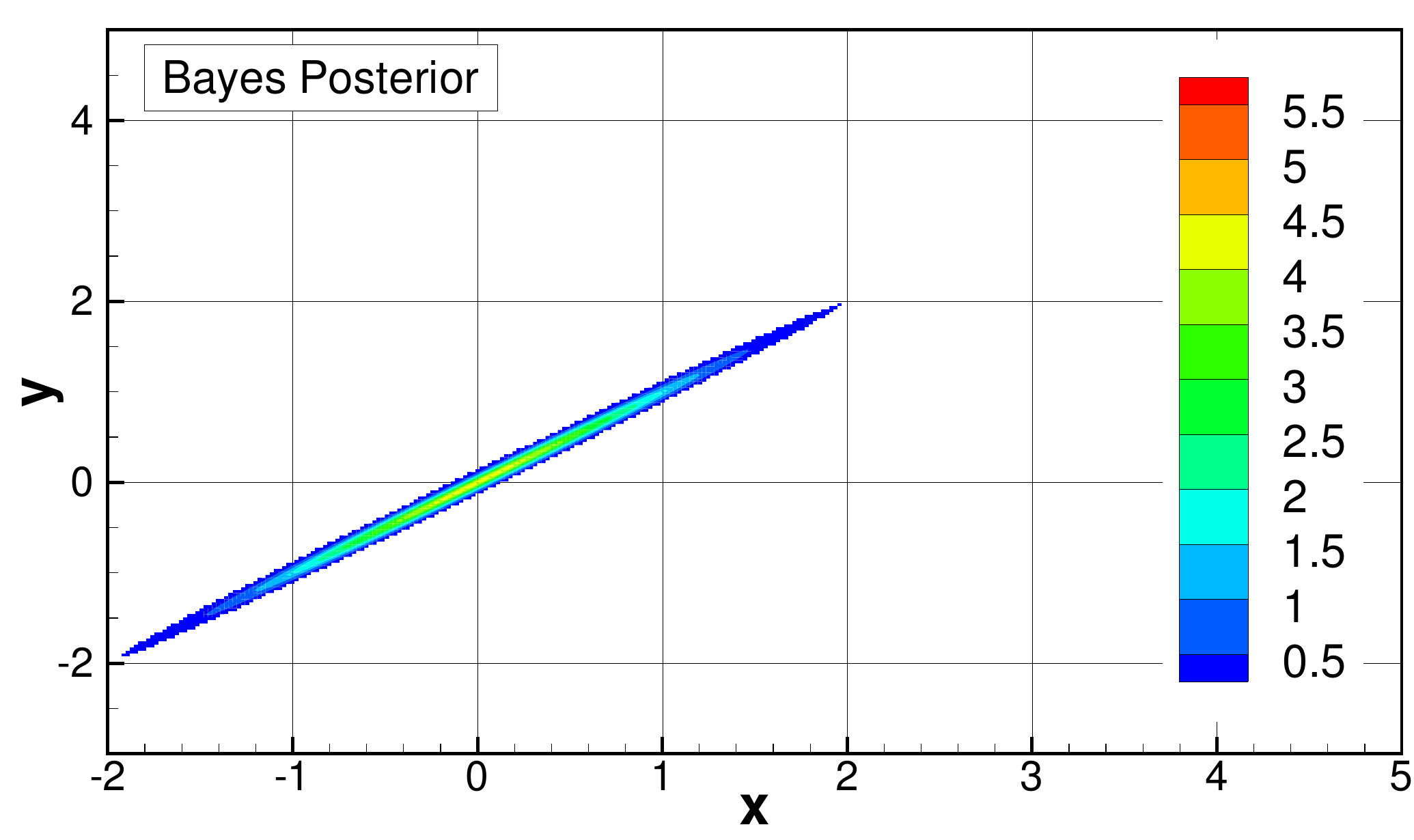}&
\includegraphics[width=0.42\textwidth,trim= 0 0 0 0]{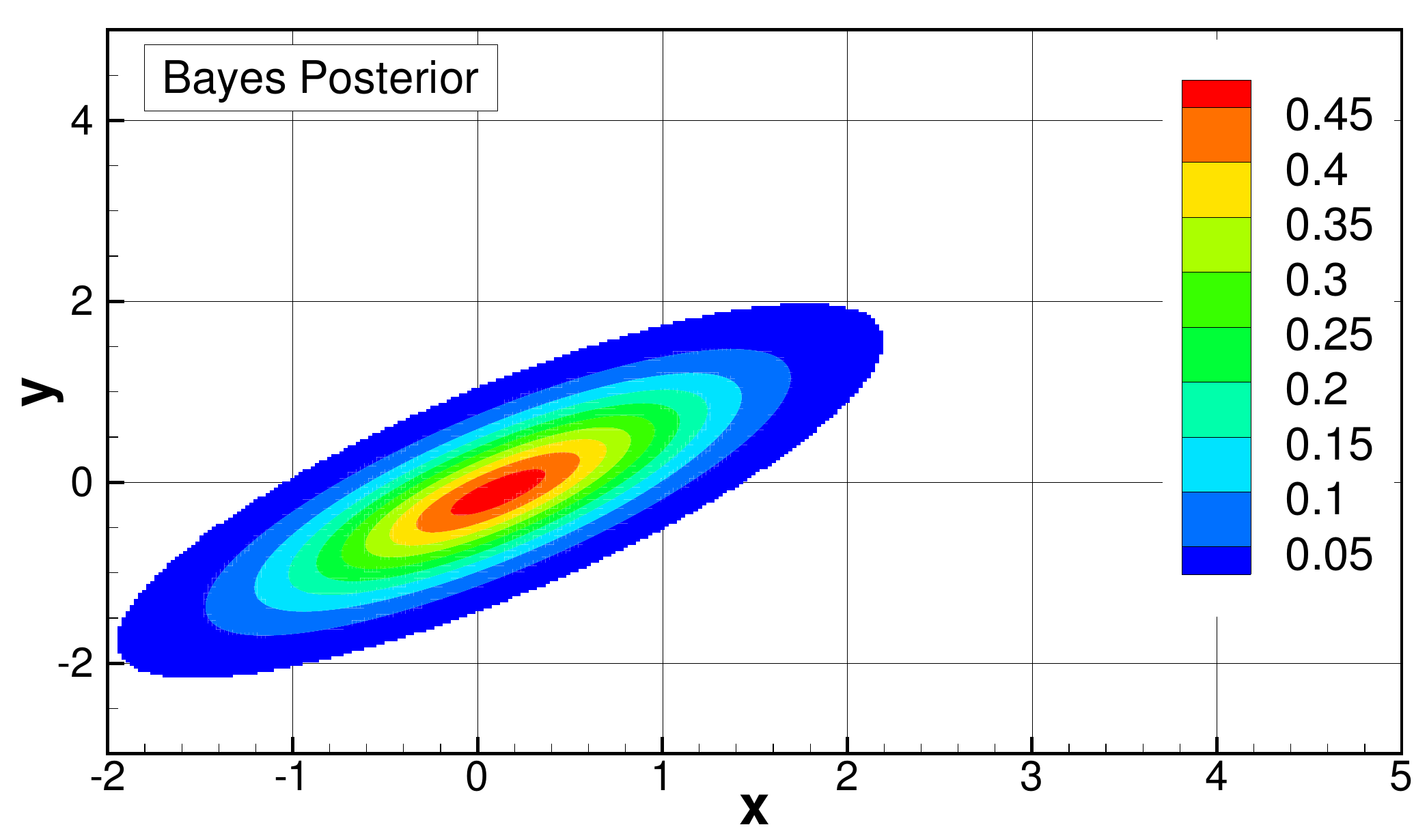}\\
\includegraphics[width=0.42\textwidth,trim= 0 0 0 0]{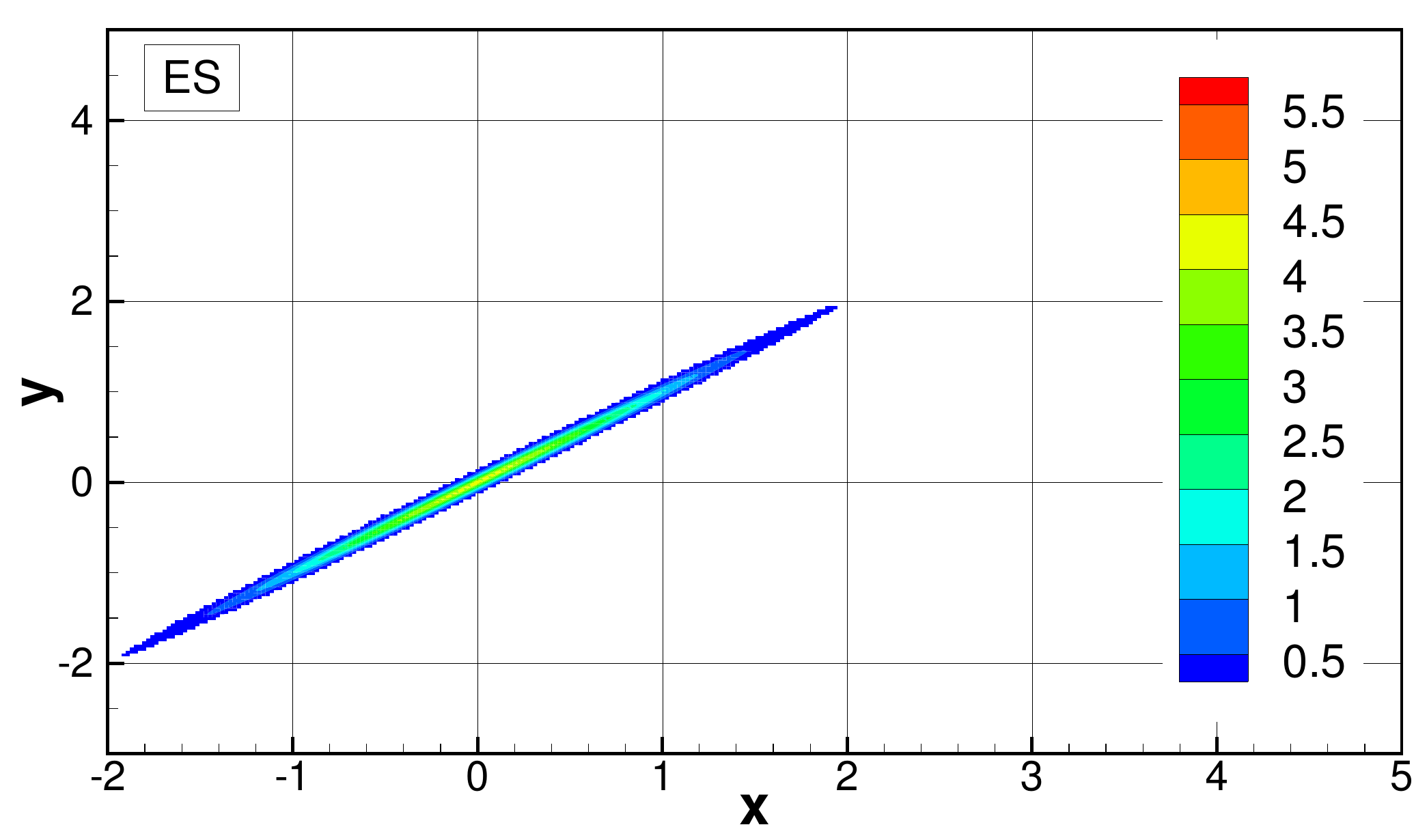}&
\includegraphics[width=0.42\textwidth,trim= 0 0 0 0]{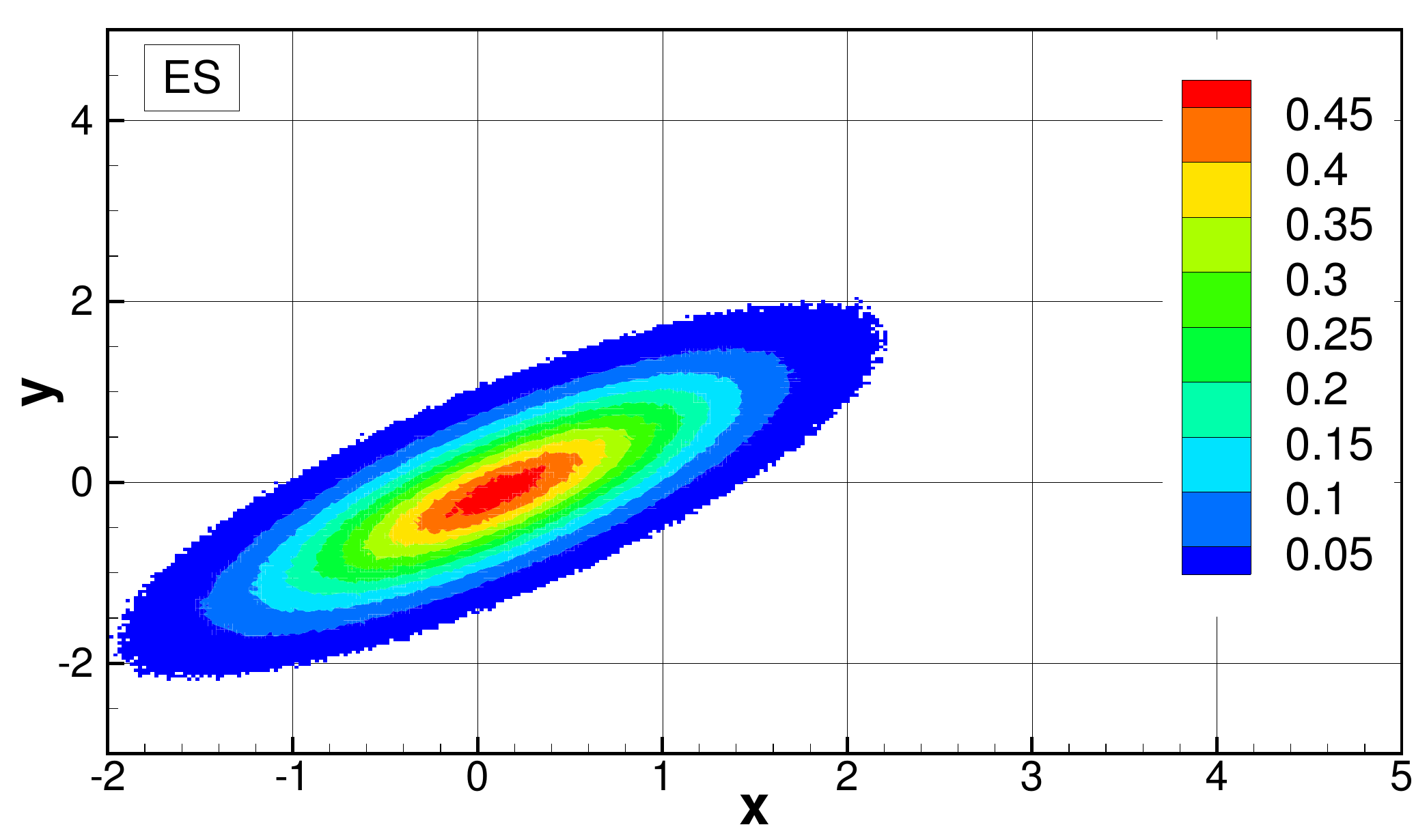}\\
\includegraphics[width=0.42\textwidth,trim= 0 0 0 0]{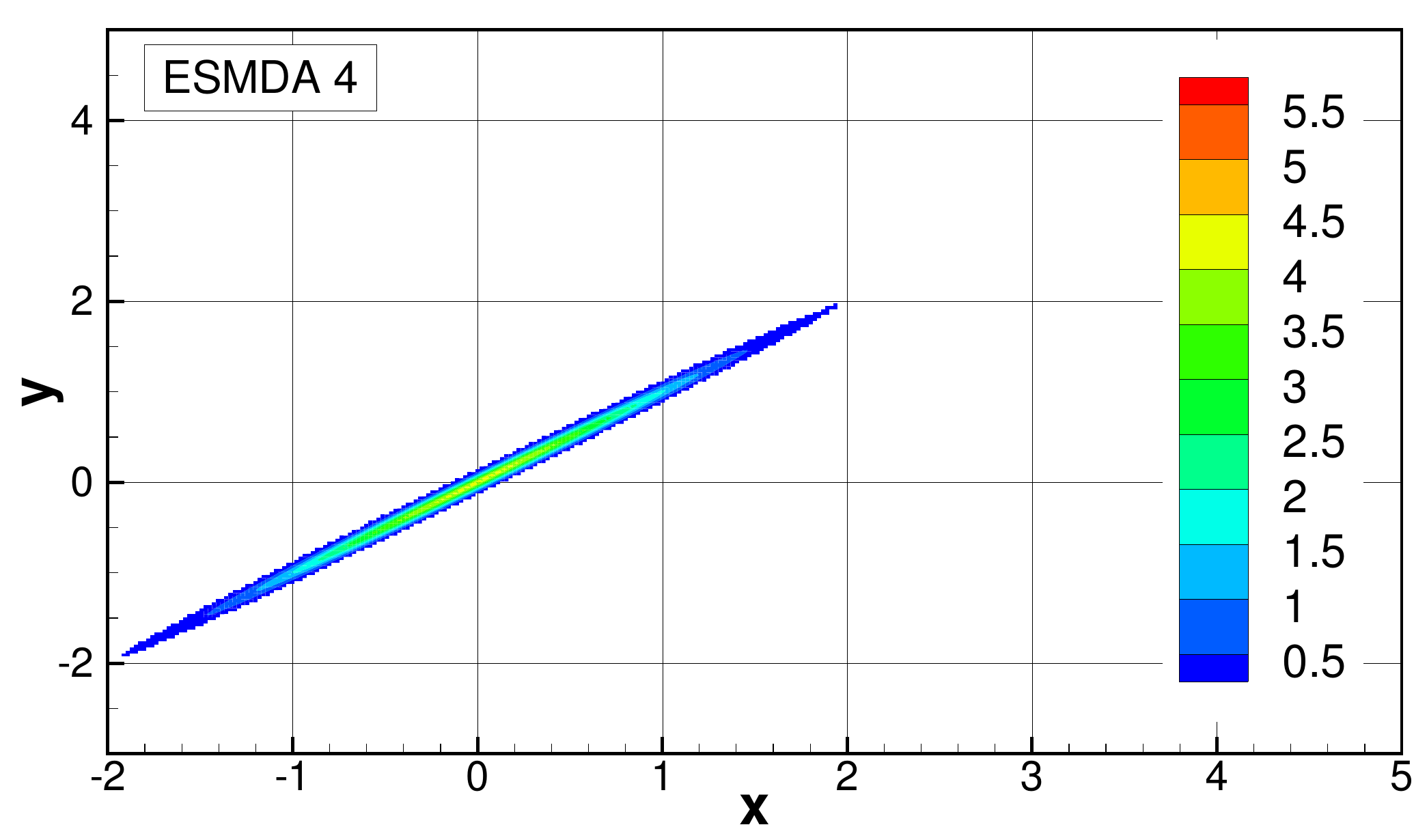}&
\includegraphics[width=0.42\textwidth,trim= 0 0 0 0]{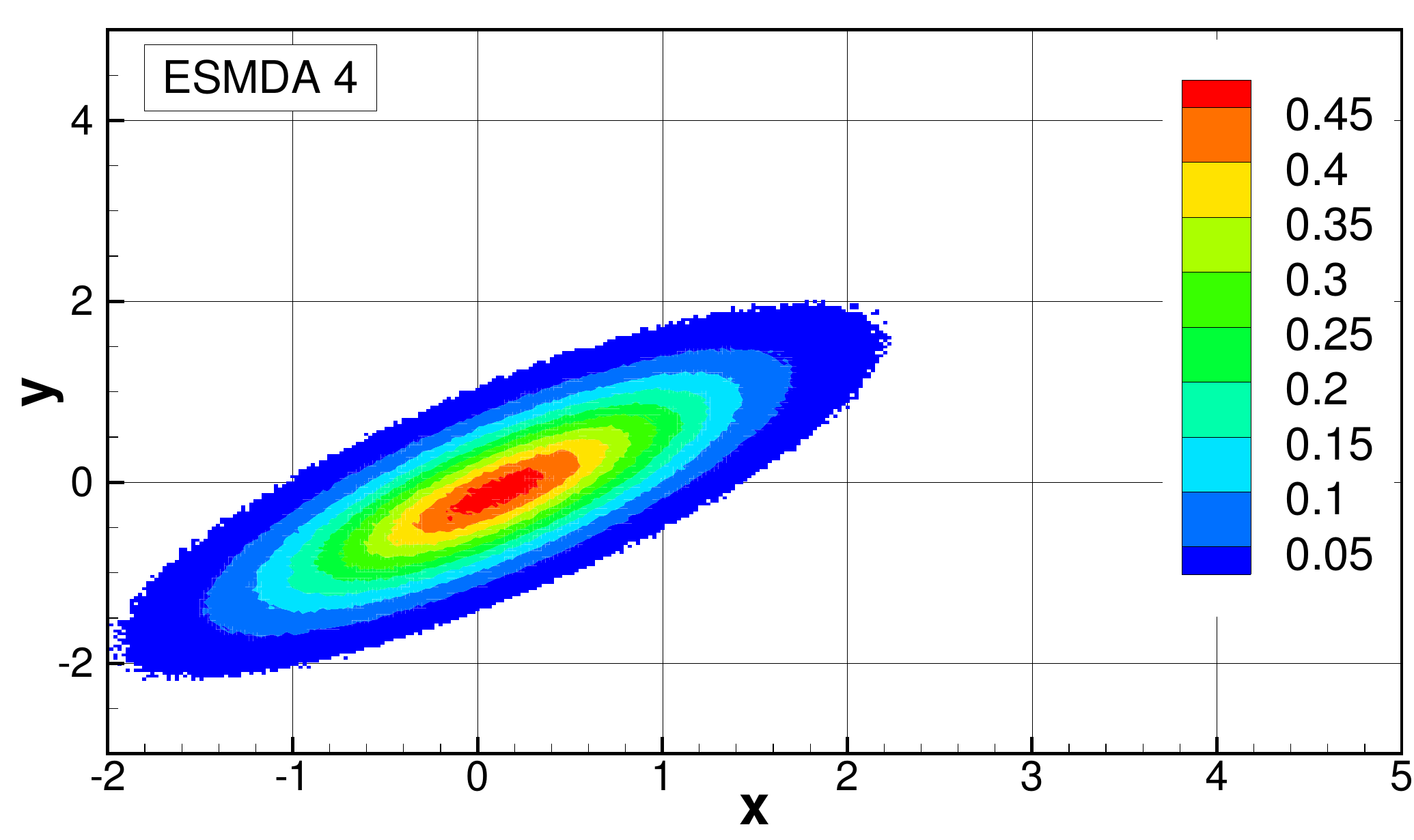}\\
\includegraphics[width=0.42\textwidth,trim= 0 0 0 0]{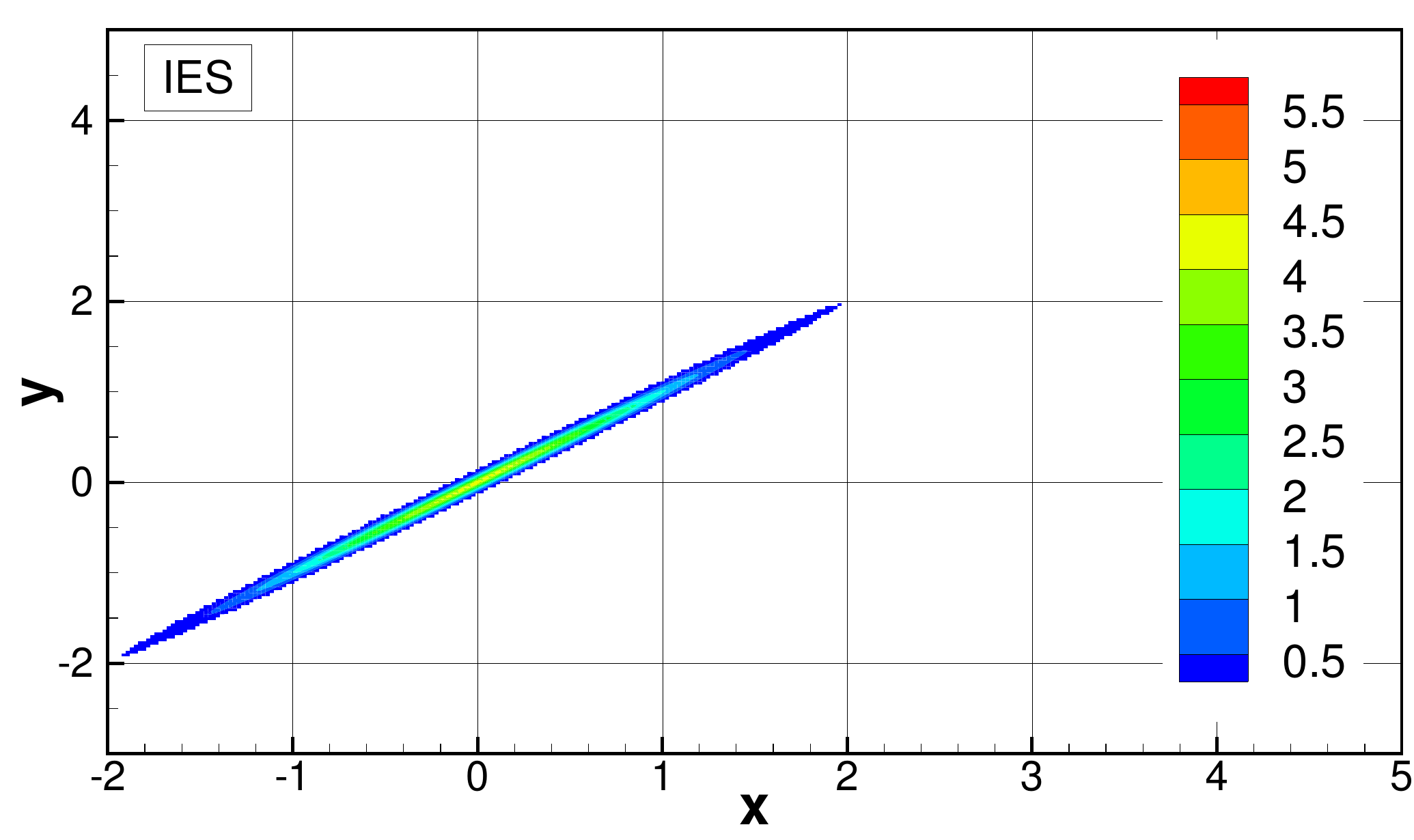}&
\includegraphics[width=0.42\textwidth,trim= 0 0 0 0]{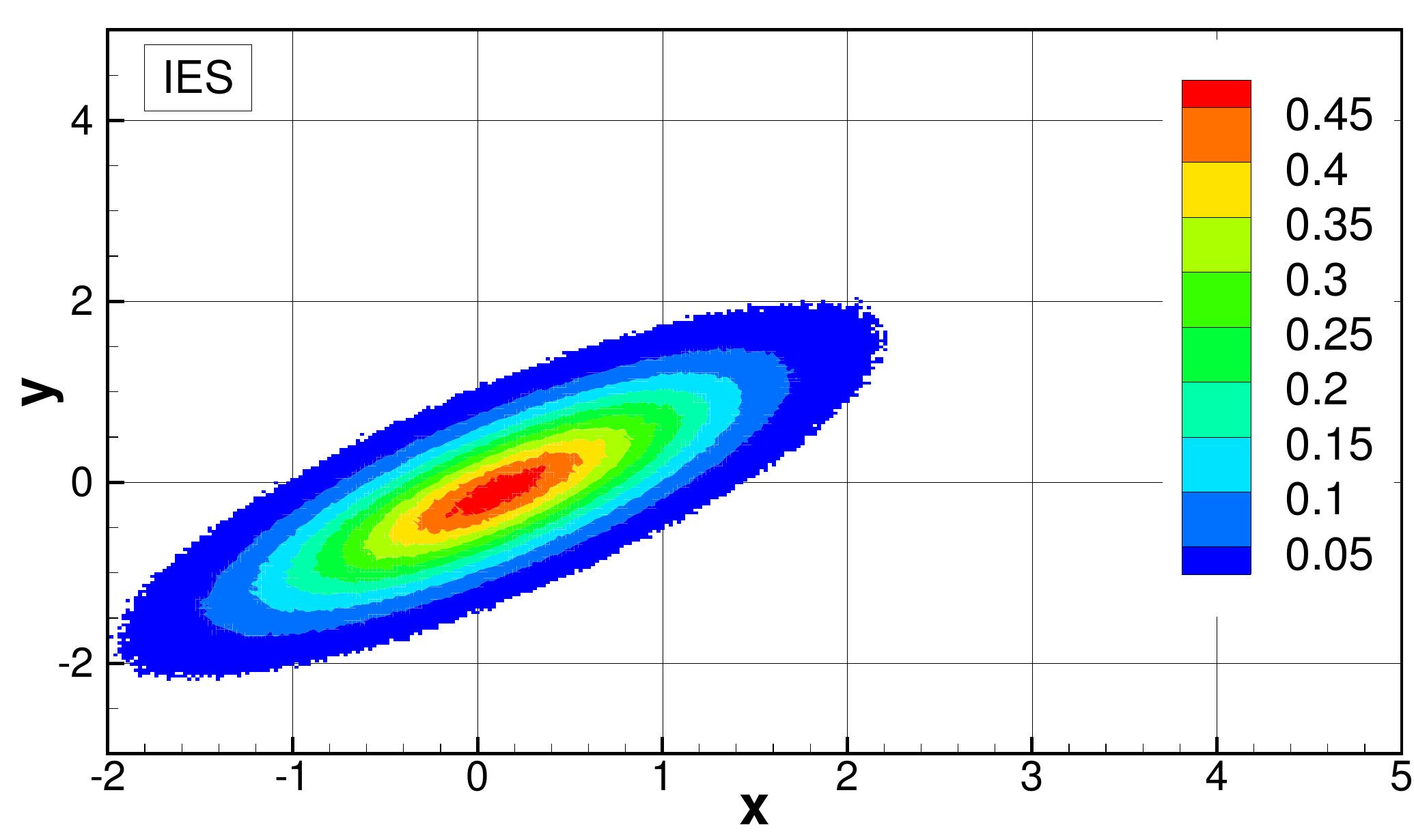}
\end{tabular}
\end{center}
\caption{
The plots show joint pdfs for the linear case. In the left column, the model error is set to zero (a small model error is retained in the final prediction for plotting purposes), while in the right
column, we include a model error with standard deviation equal to 0.5. The two upper rows are the analytical prior and posterior,
while the three lower rows show the results from ES, ESMDA with four steps, and IES.
 \label{fig:pdfL}}
\end{figure*}

\begin{figure*}[t]
\begin{center}
\tabcolsep=0.0pt
\begin{tabular}{cc}
\includegraphics[width=0.5\textwidth,trim= 0 0 0 0]{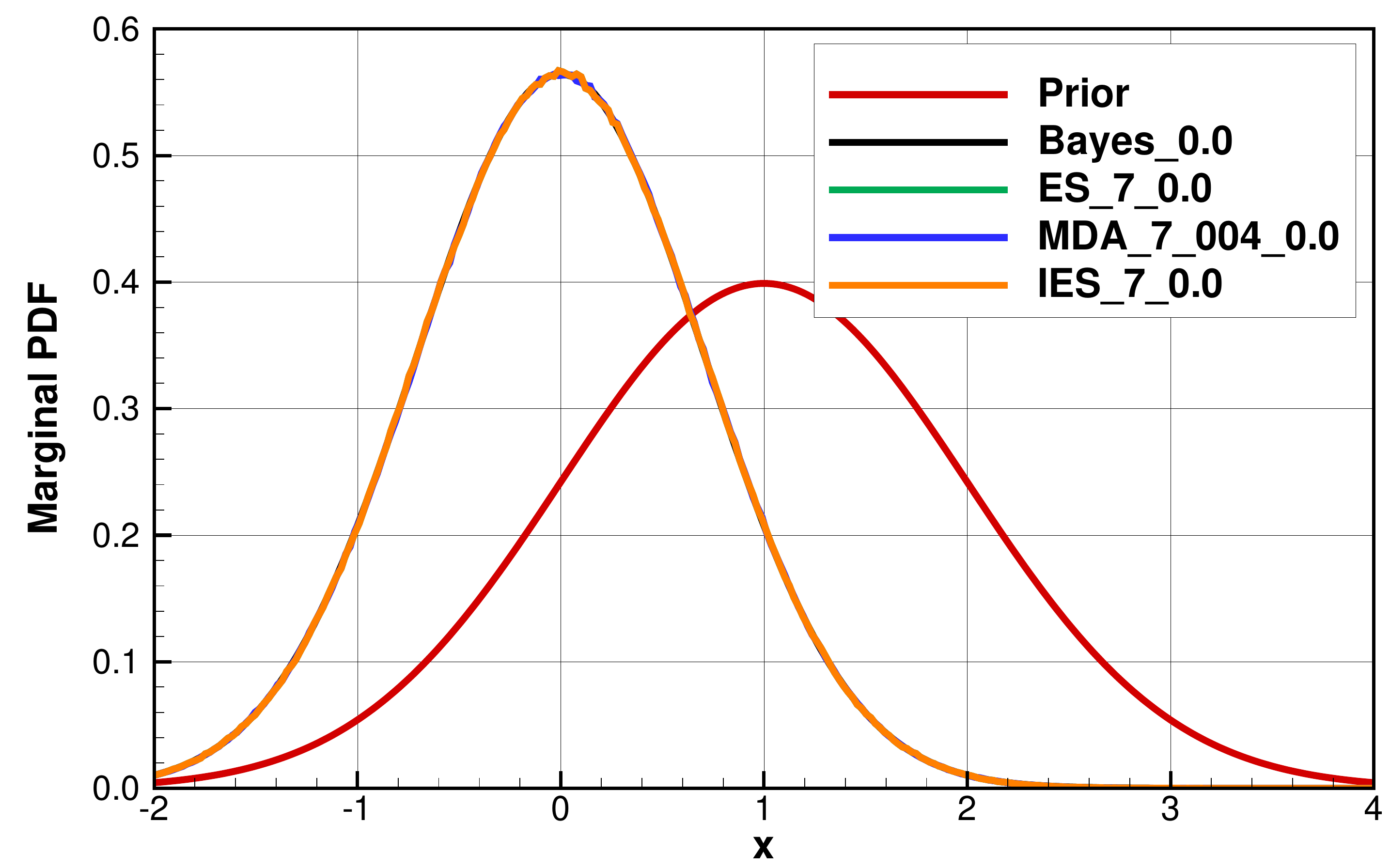}&
\includegraphics[width=0.5\textwidth,trim= 0 0 0 0]{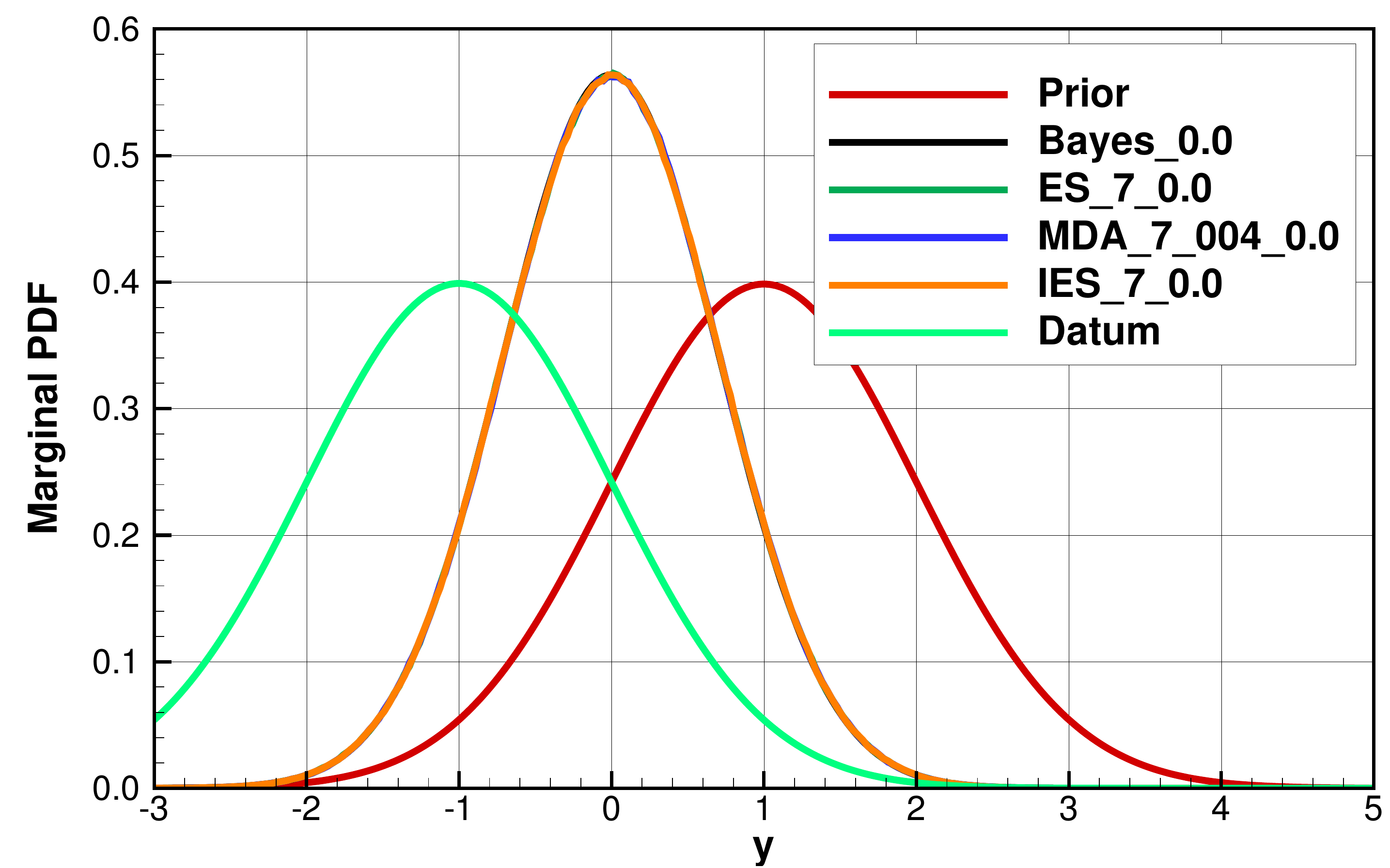}\\
\includegraphics[width=0.5\textwidth,trim= 0 0 0 0]{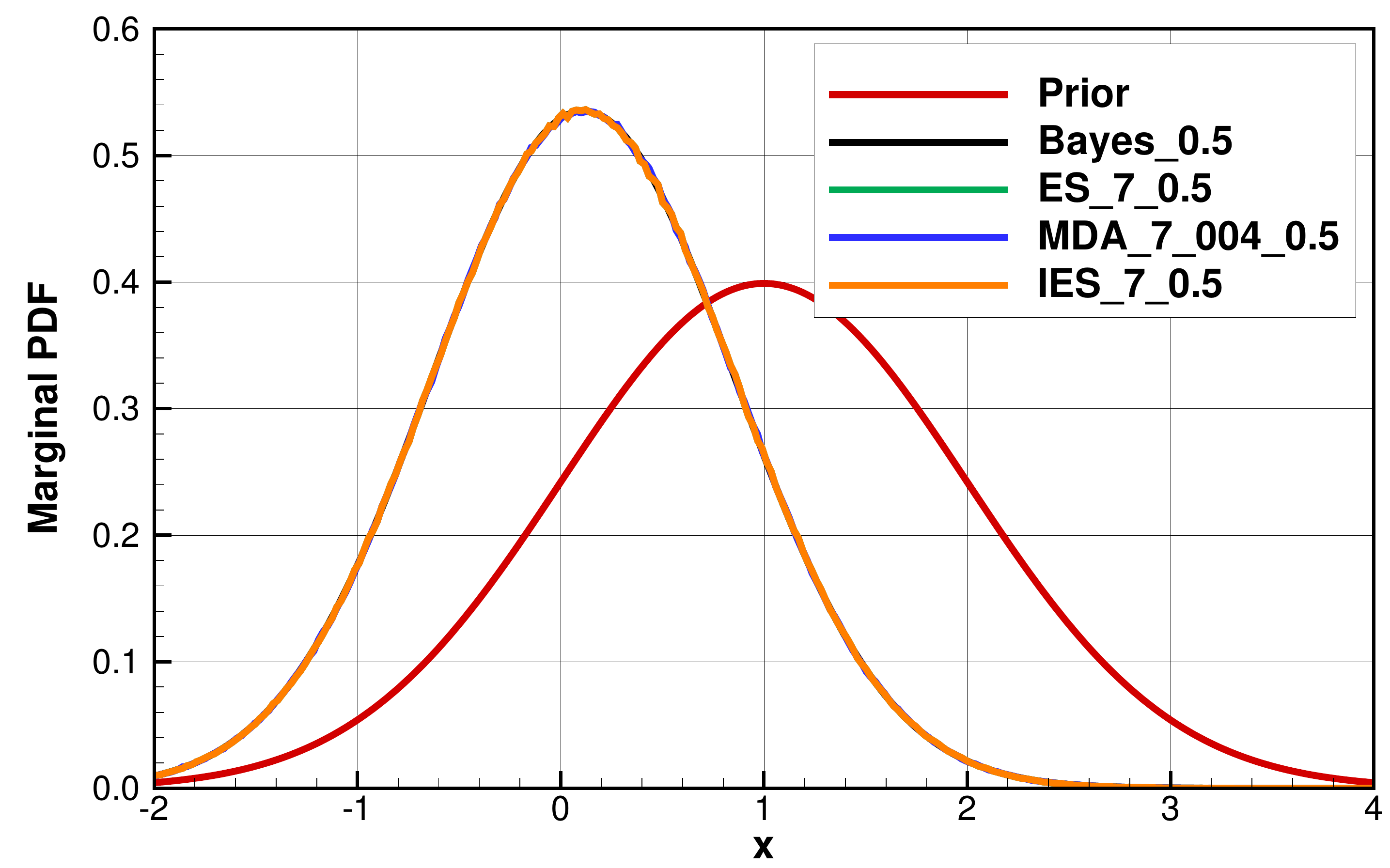}&
\includegraphics[width=0.5\textwidth,trim= 0 0 0 0]{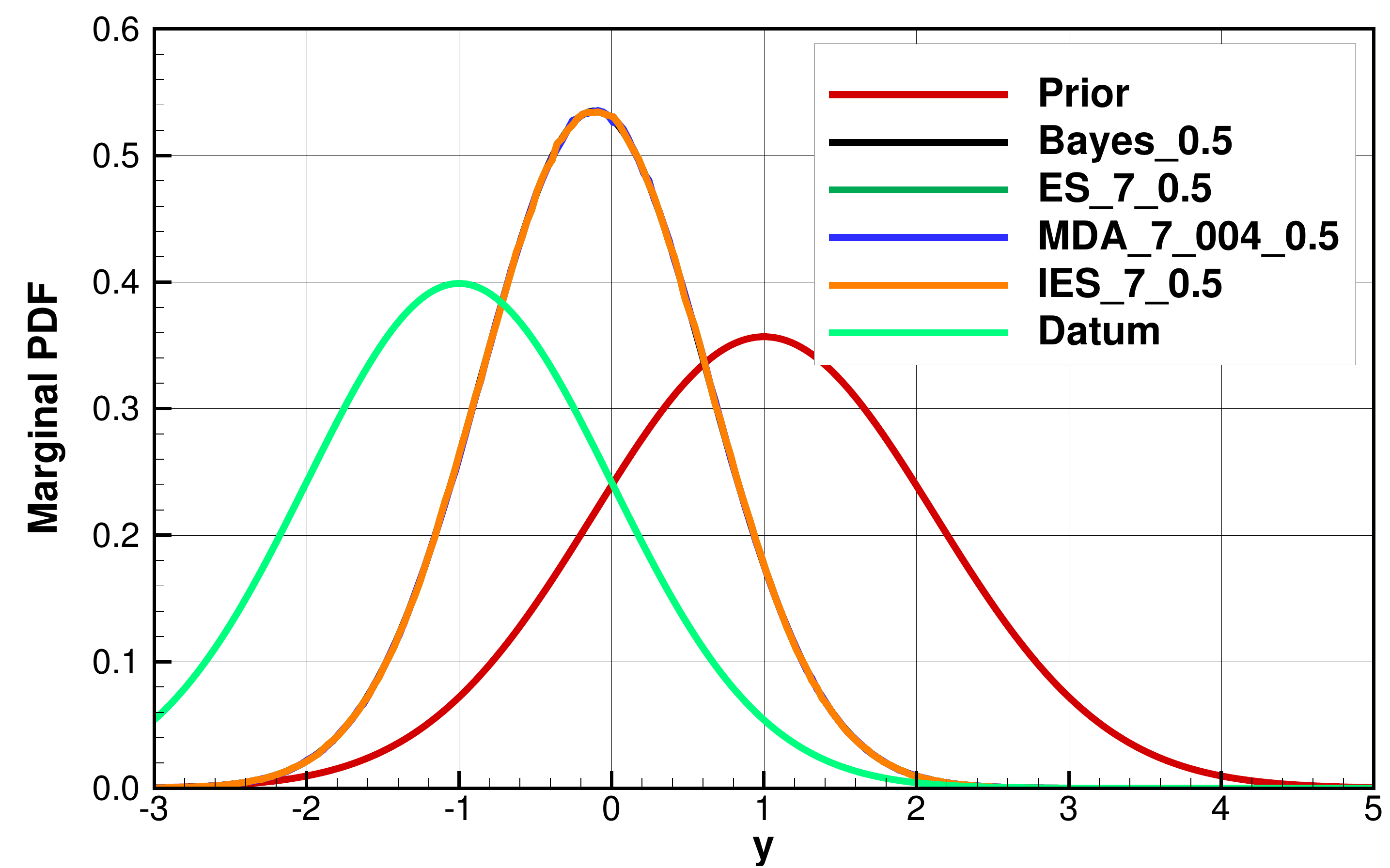}
\end{tabular}
\end{center}
\caption{The plots show the marginal pdfs for x in the left column and the marginal pdfs for y in the right column for the linear case corresponding to Figure~\protect\ref{fig:pdfL}.
The upper row is the case with zero model error, while the lower row shows the result when we include a model error with standard deviation equal to 0.5.  In all the plots, the 
legends, e.g., MDA$\_7\_004\_0.5$ denote ESMDA with $10^7$ ensemble members, 4 MDA steps, and a model error with variance equal to 0.5.  Note that, in the linear case, all the methods give results that are identical to
the Bayesian posterior, and the posterior pdfs are indistinguishable.
 \label{fig:margL}}
\end{figure*}

\begin{figure}[t]
\begin{center}
\tabcolsep=0.0pt
\begin{tabular}{c}
\includegraphics[width=1.0\columnwidth,trim= 0 0 0 0]{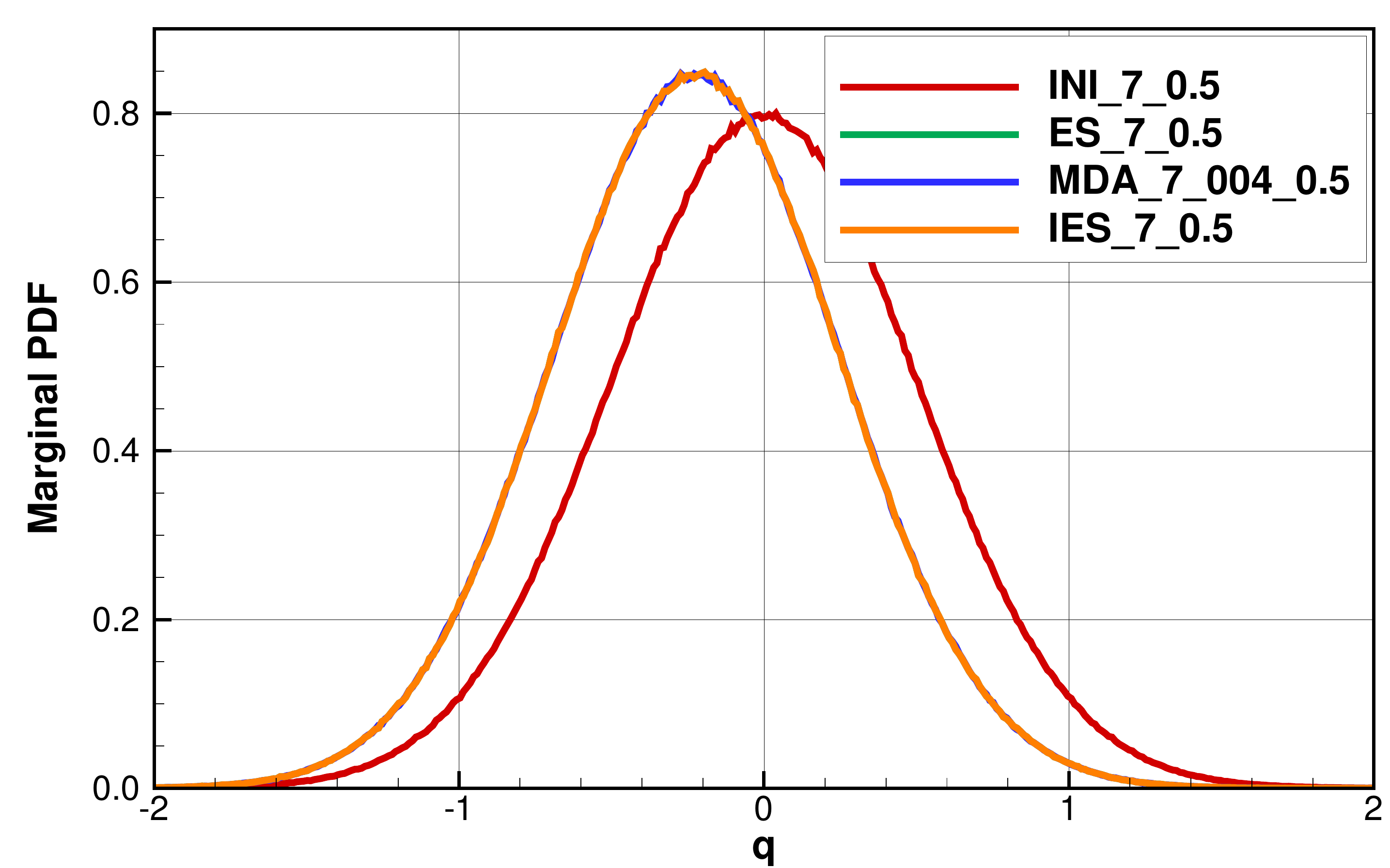}
\end{tabular}
\end{center}
\caption{Model error distributions for the linear case. It is not possible to distinguish between the pdfs for the estimated model errors from the different methods.
 \label{fig:margQL}}
\end{figure}


\begin{figure*}[p]
\begin{center}
\begin{tabular}{cc}
\includegraphics[width=0.42\textwidth,trim= 0 0 0 0]{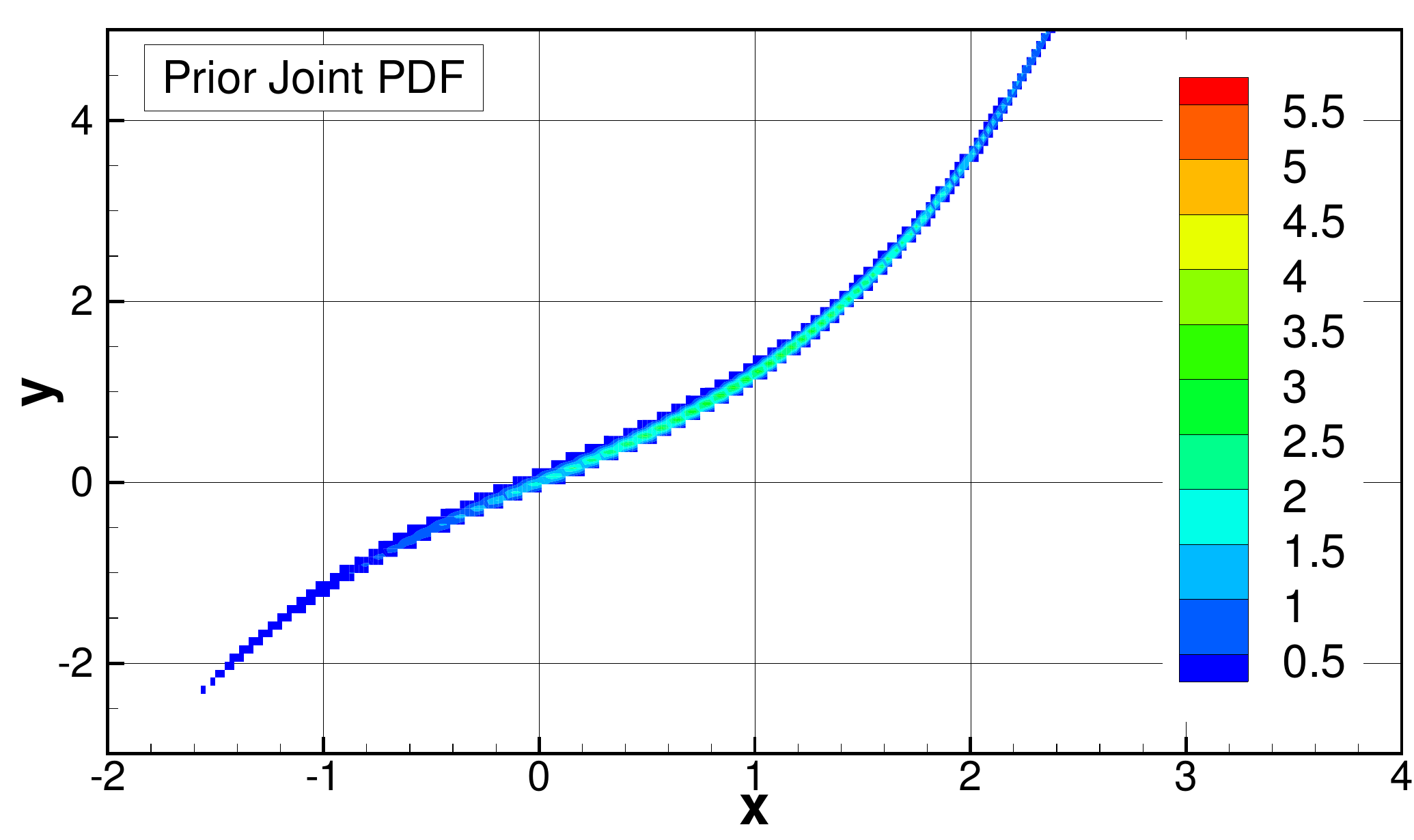}&
\includegraphics[width=0.42\textwidth,trim= 0 0 0 0]{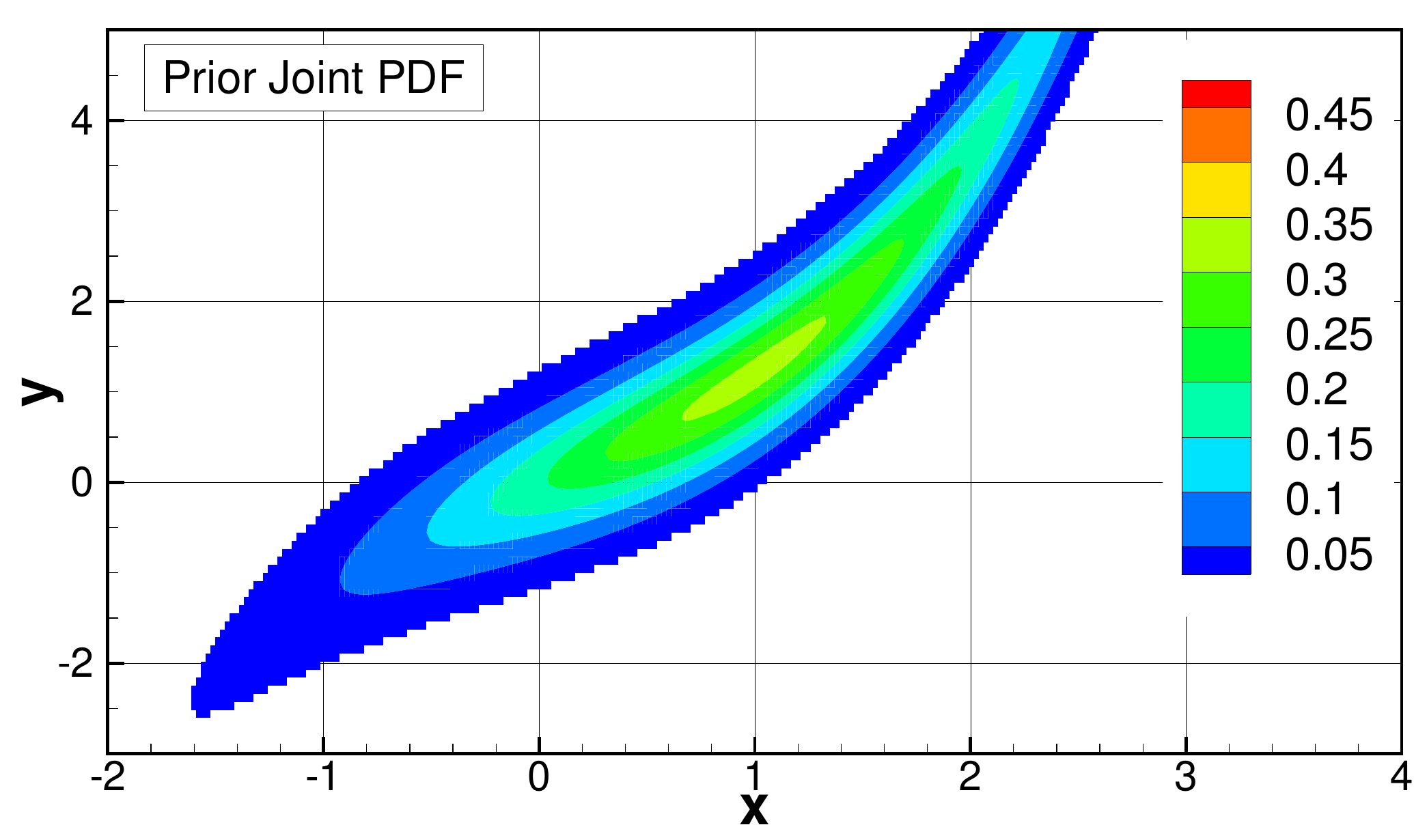}\\
\includegraphics[width=0.42\textwidth,trim= 0 0 0 0]{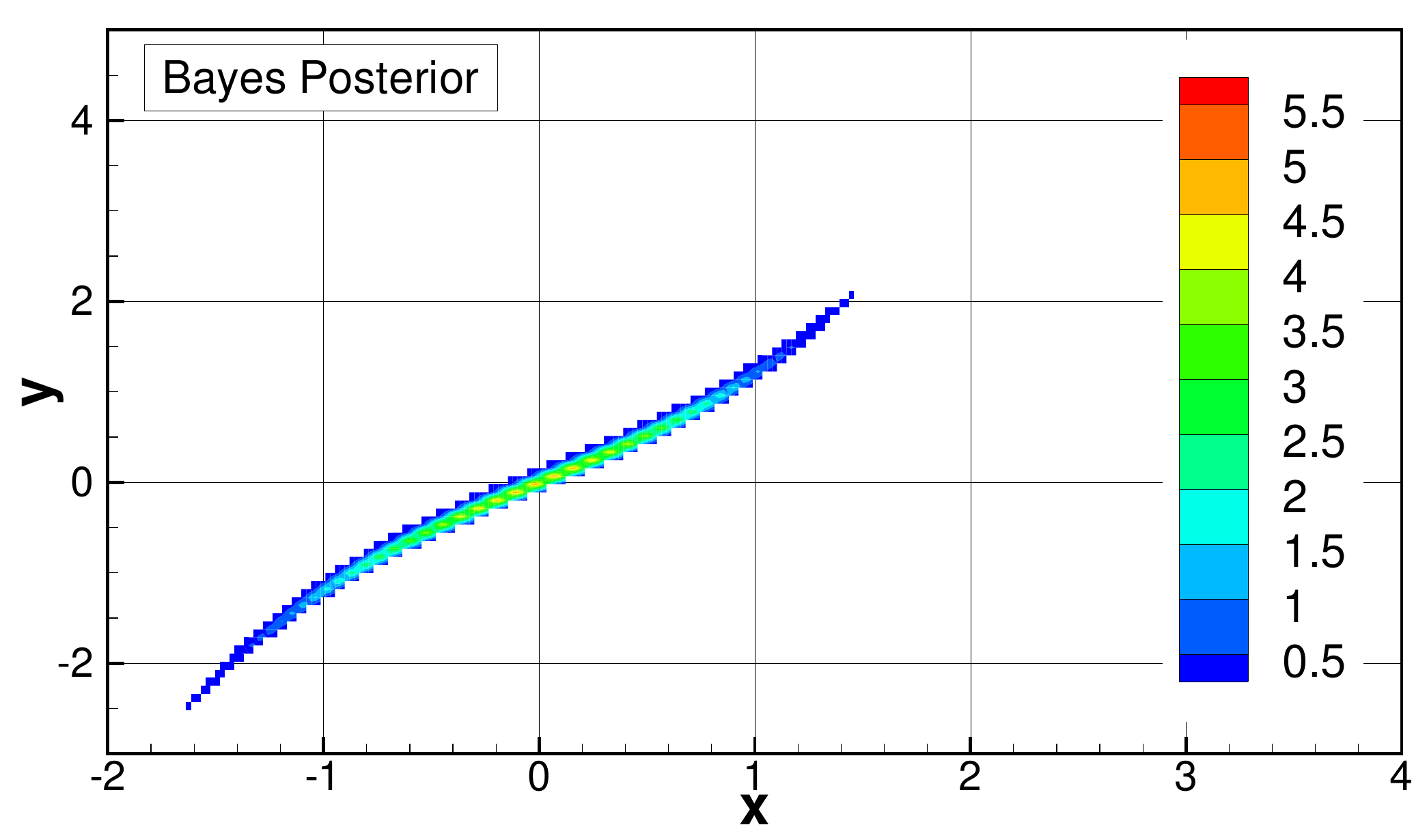}&
\includegraphics[width=0.42\textwidth,trim= 0 0 0 0]{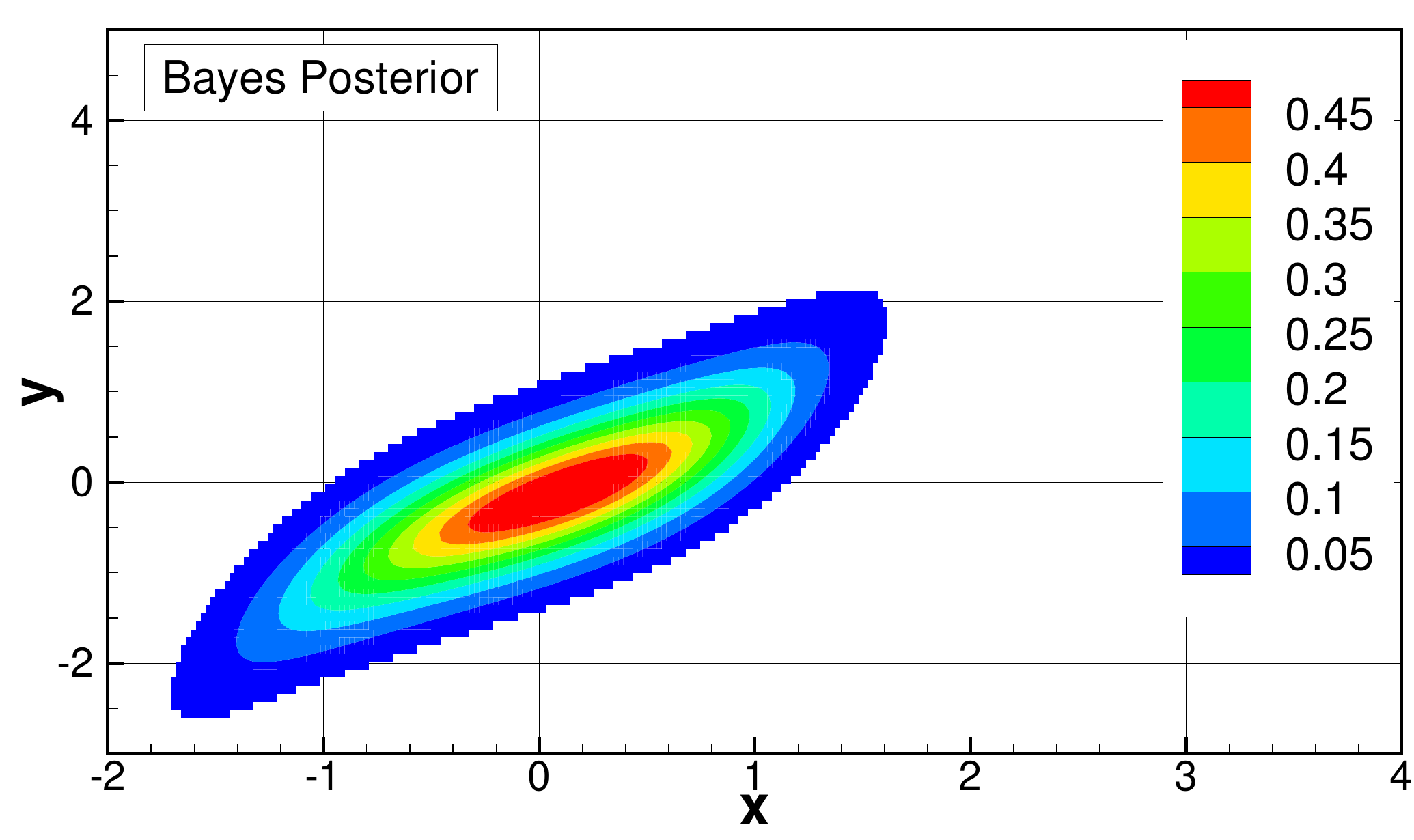}\\
\includegraphics[width=0.42\textwidth,trim= 0 0 0 0]{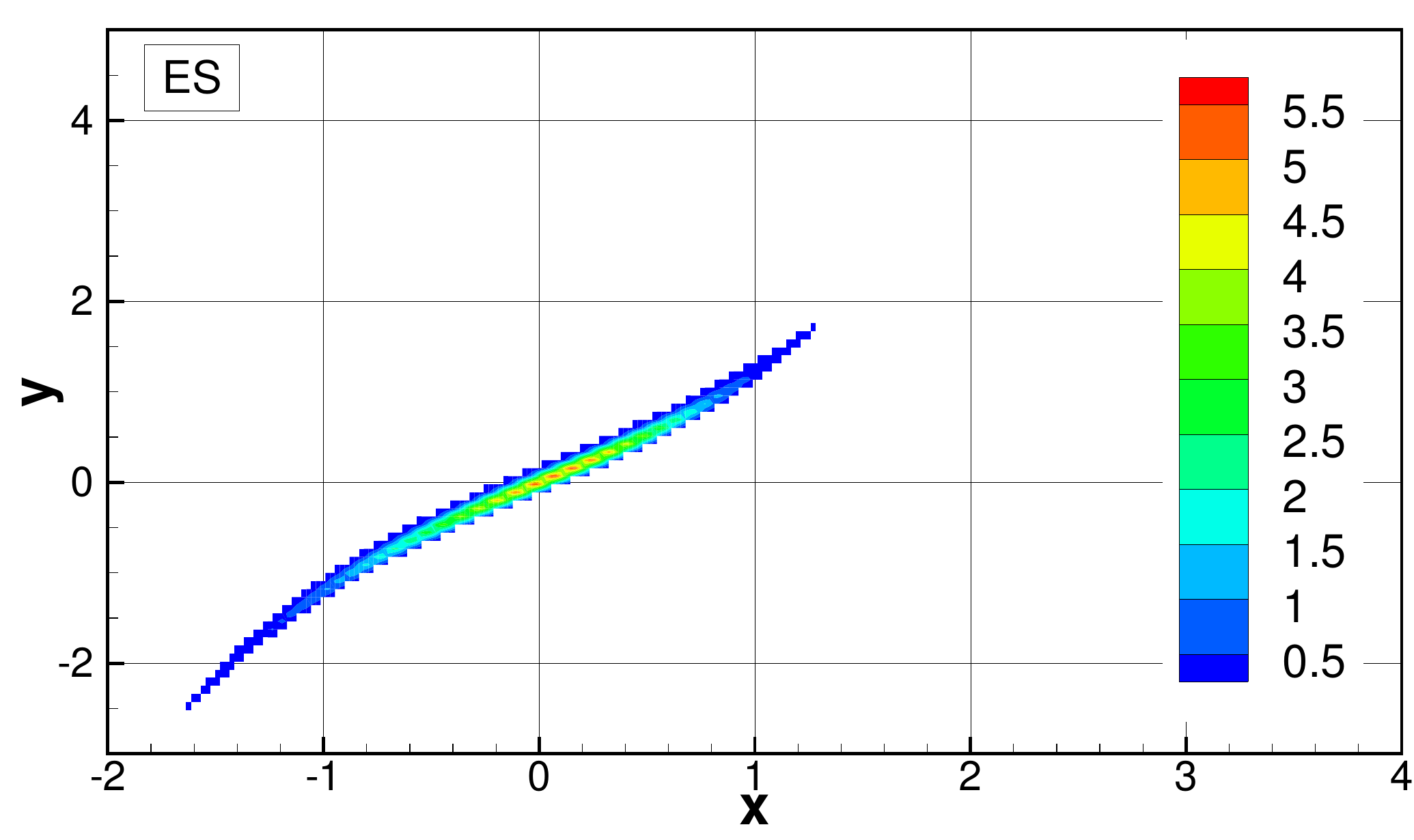}&
\includegraphics[width=0.42\textwidth,trim= 0 0 0 0]{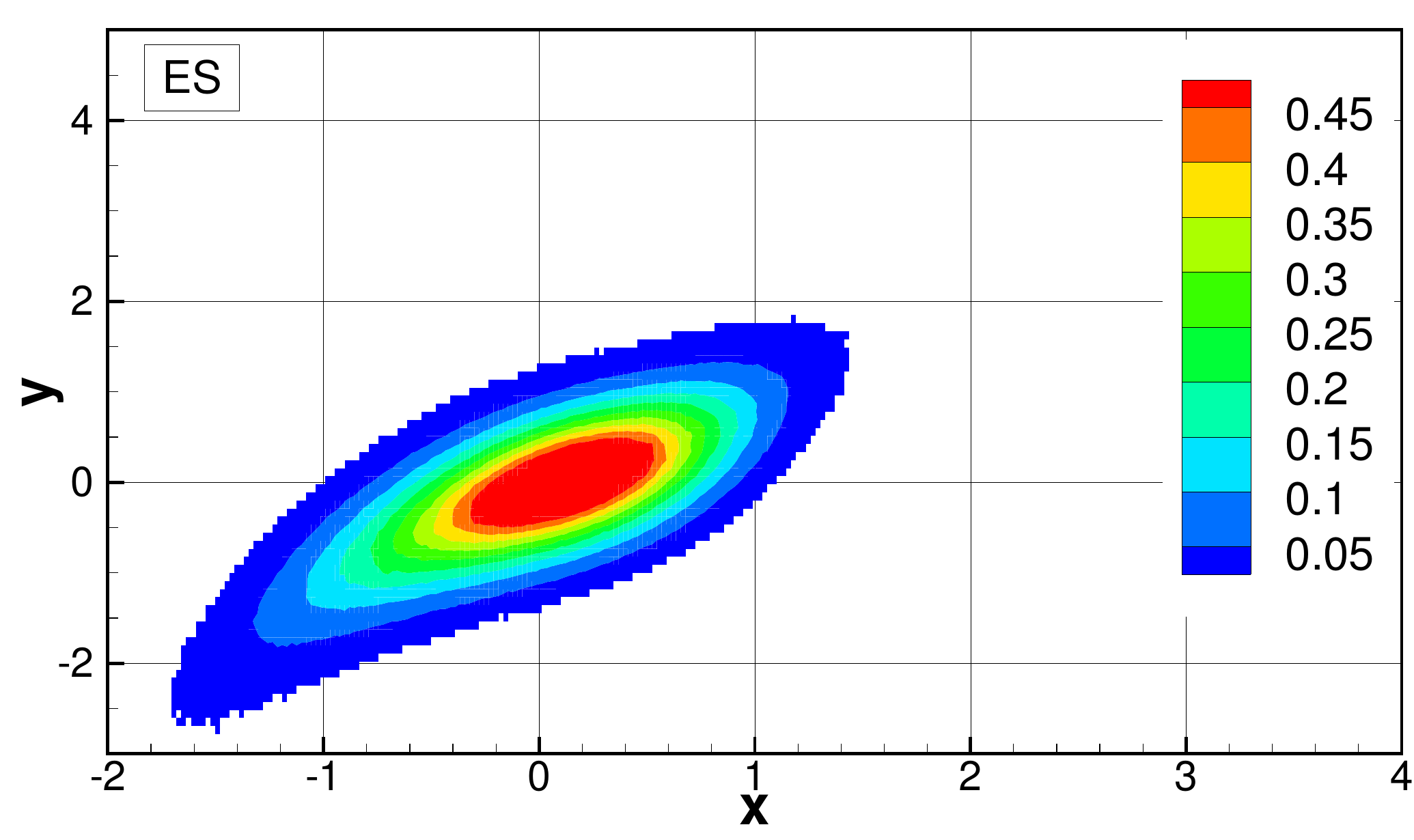}\\
\includegraphics[width=0.42\textwidth,trim= 0 0 0 0]{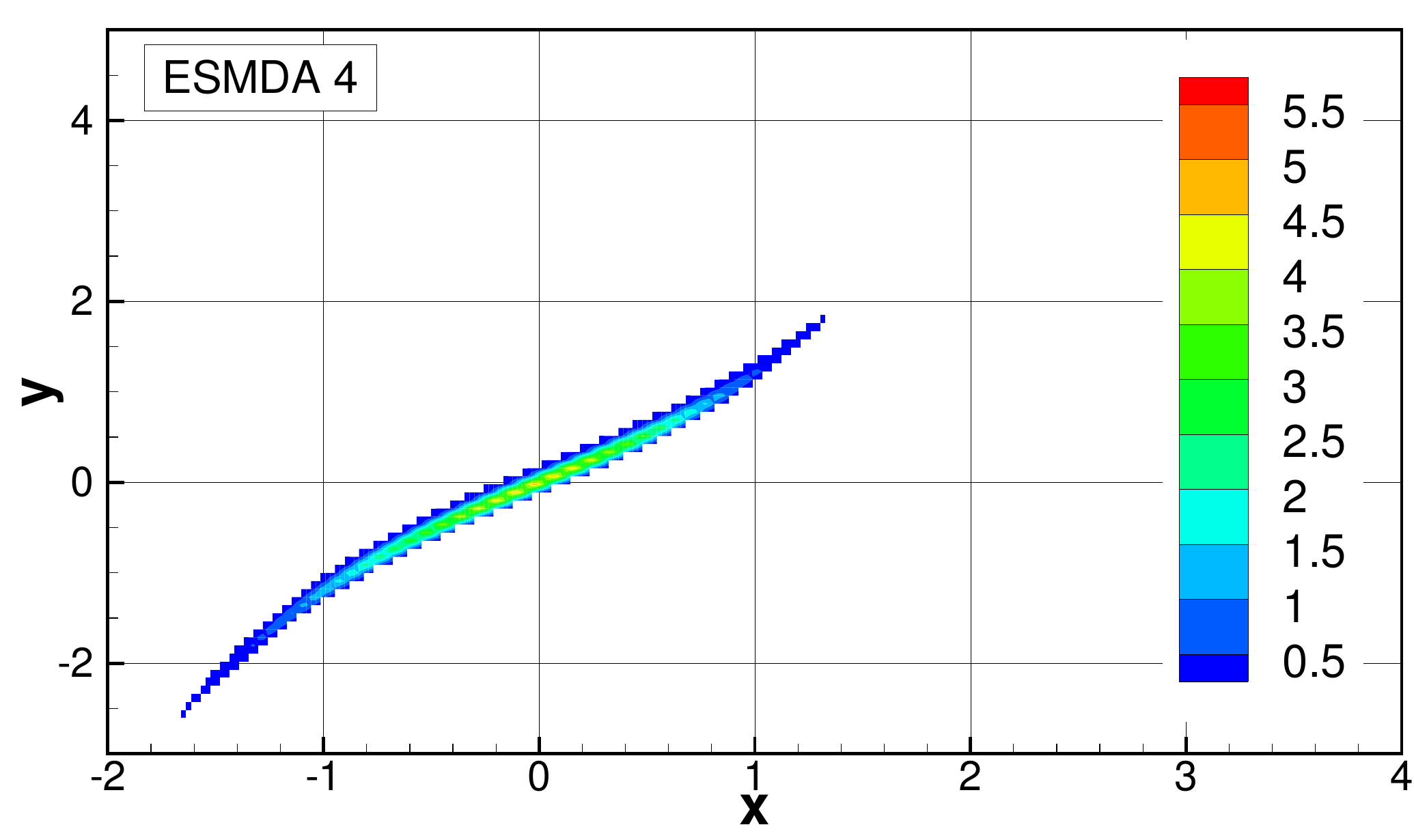}&
\includegraphics[width=0.42\textwidth,trim= 0 0 0 0]{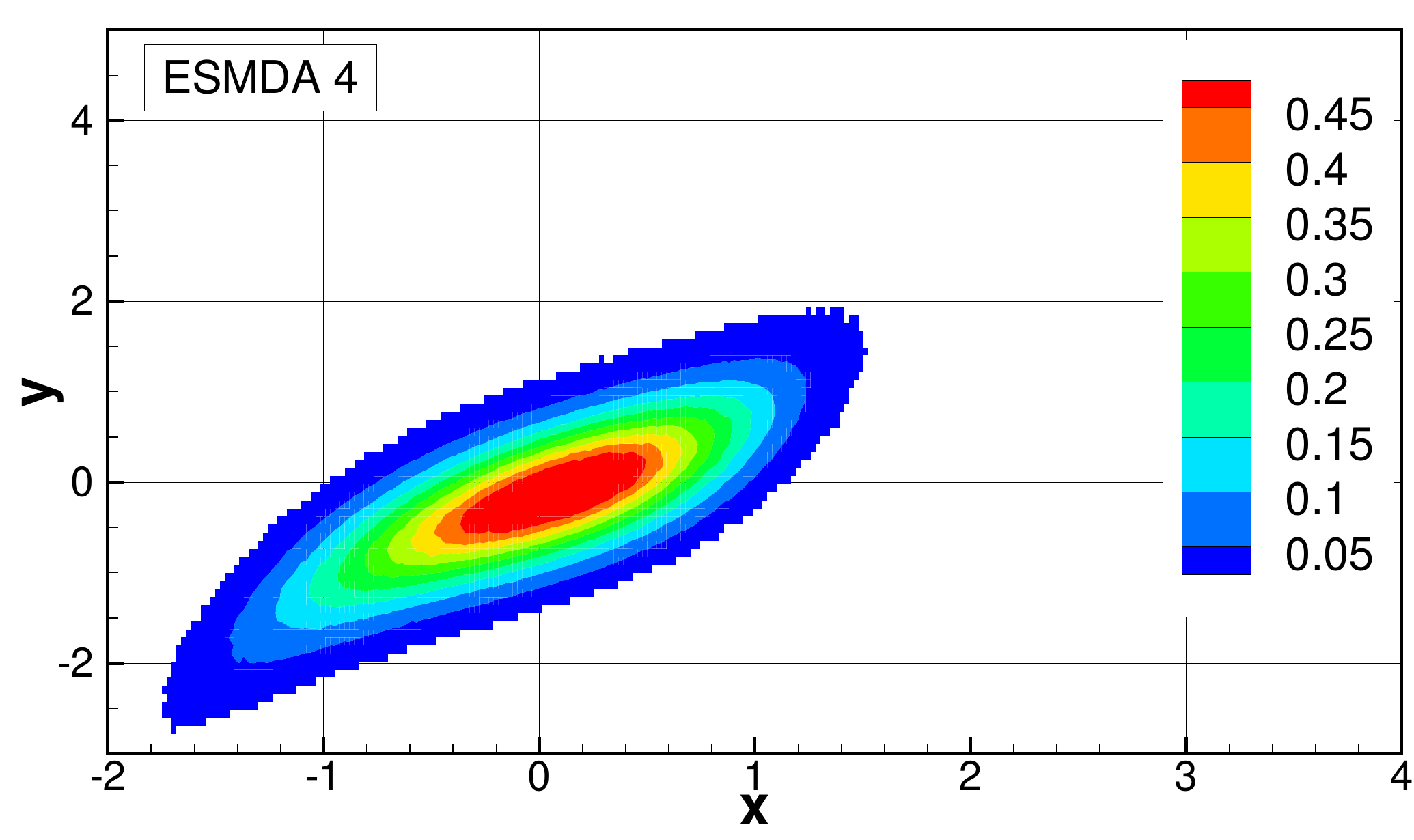}\\
\includegraphics[width=0.42\textwidth,trim= 0 0 0 0]{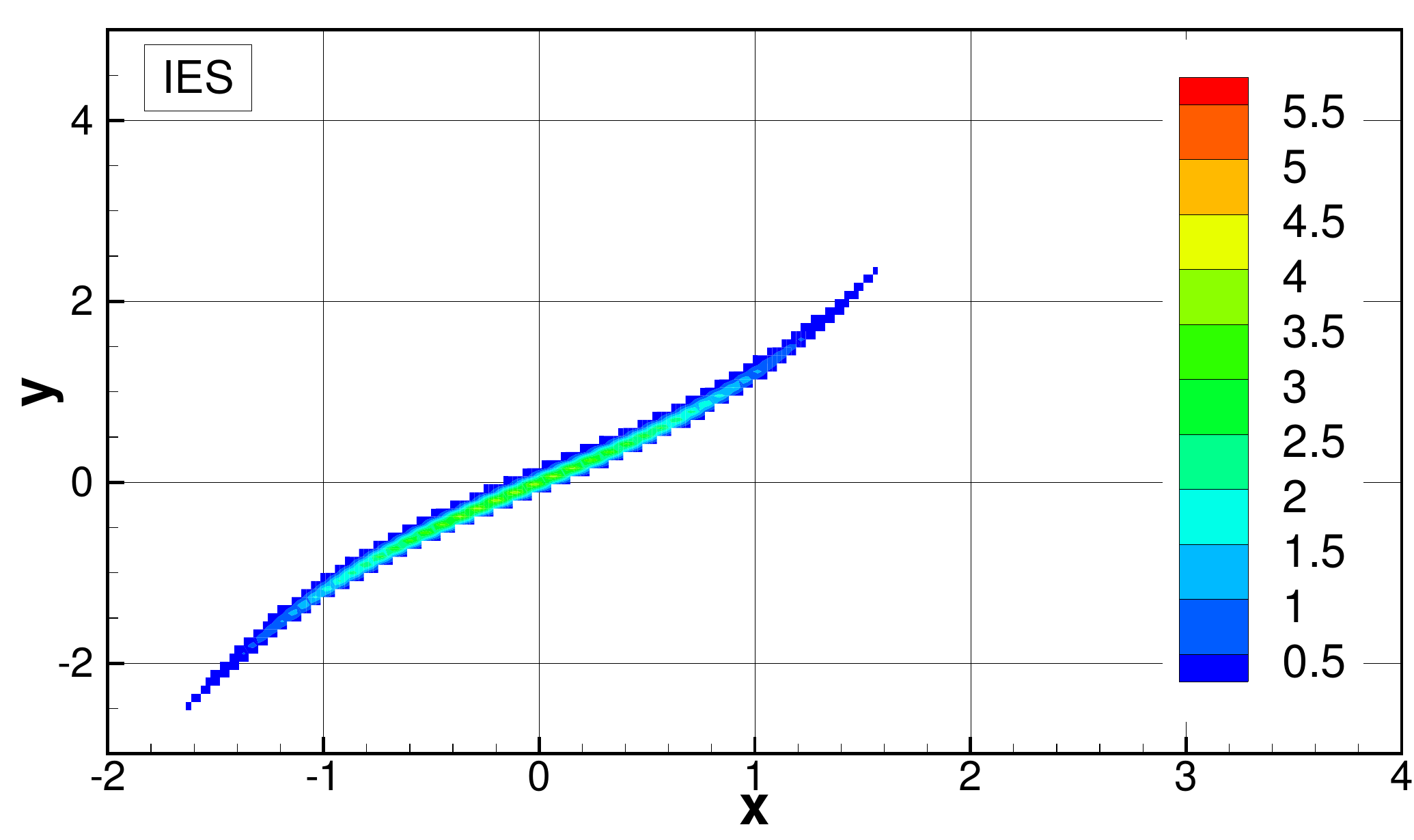}&
\includegraphics[width=0.42\textwidth,trim= 0 0 0 0]{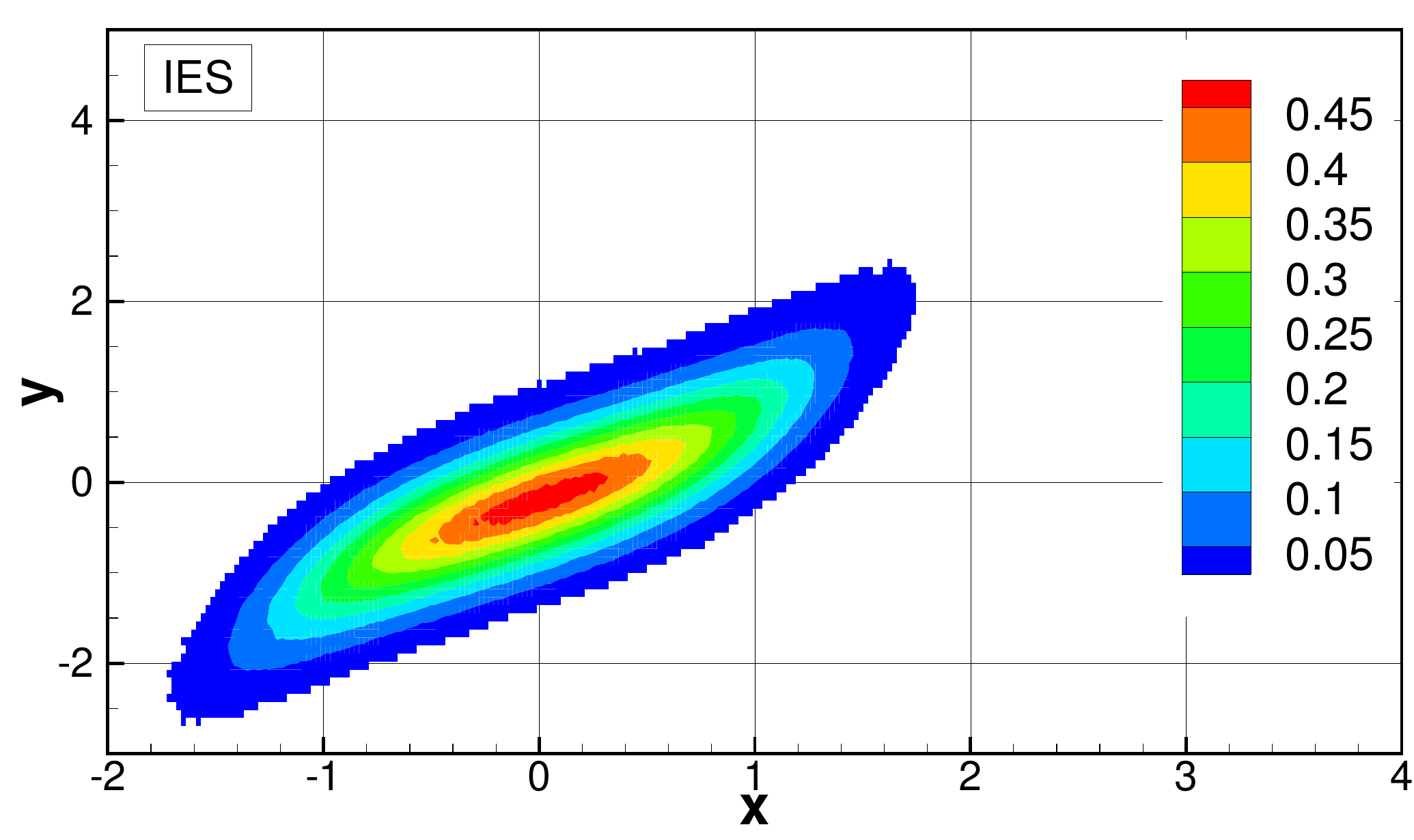}
\end{tabular}
\end{center}
\caption{The plots show joint pdfs for the nonlinear case.  In the left column, the model error is set to zero, while in the right
column, we include a model error with standard deviation equal to 0.5. The two upper rows are the analytical prior and posterior,
while the three lower rows show results from ES, ESMDA with four steps, and IES.
 \label{fig:pdfN}}
\end{figure*}

\begin{figure*}[t]
\begin{center}
\tabcolsep=0.0pt
\begin{tabular}{cc}
\includegraphics[width=0.5\textwidth,trim= 0 0 0 0]{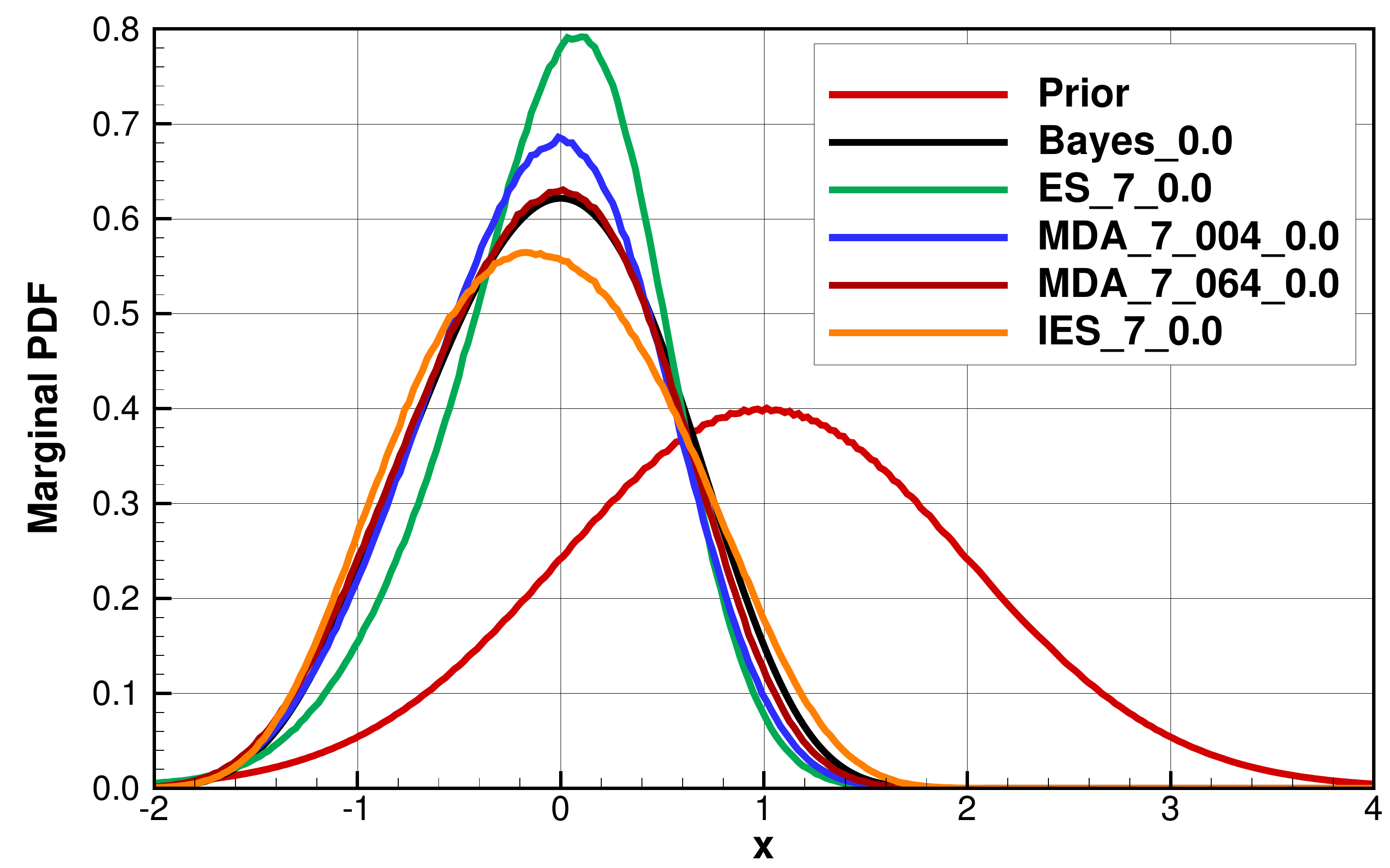}&
\includegraphics[width=0.5\textwidth,trim= 0 0 0 0]{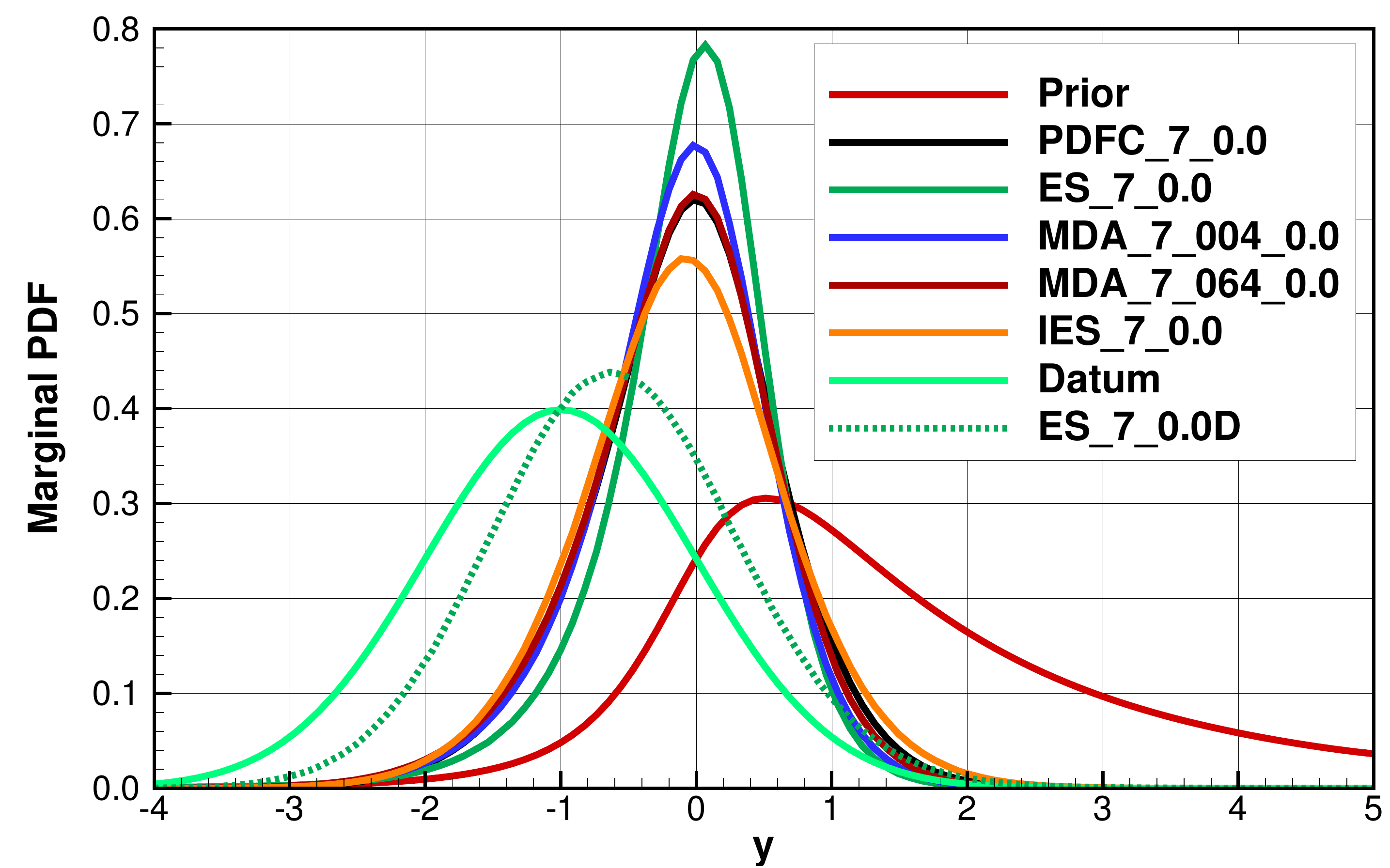}\\
\includegraphics[width=0.5\textwidth,trim= 0 0 0 0]{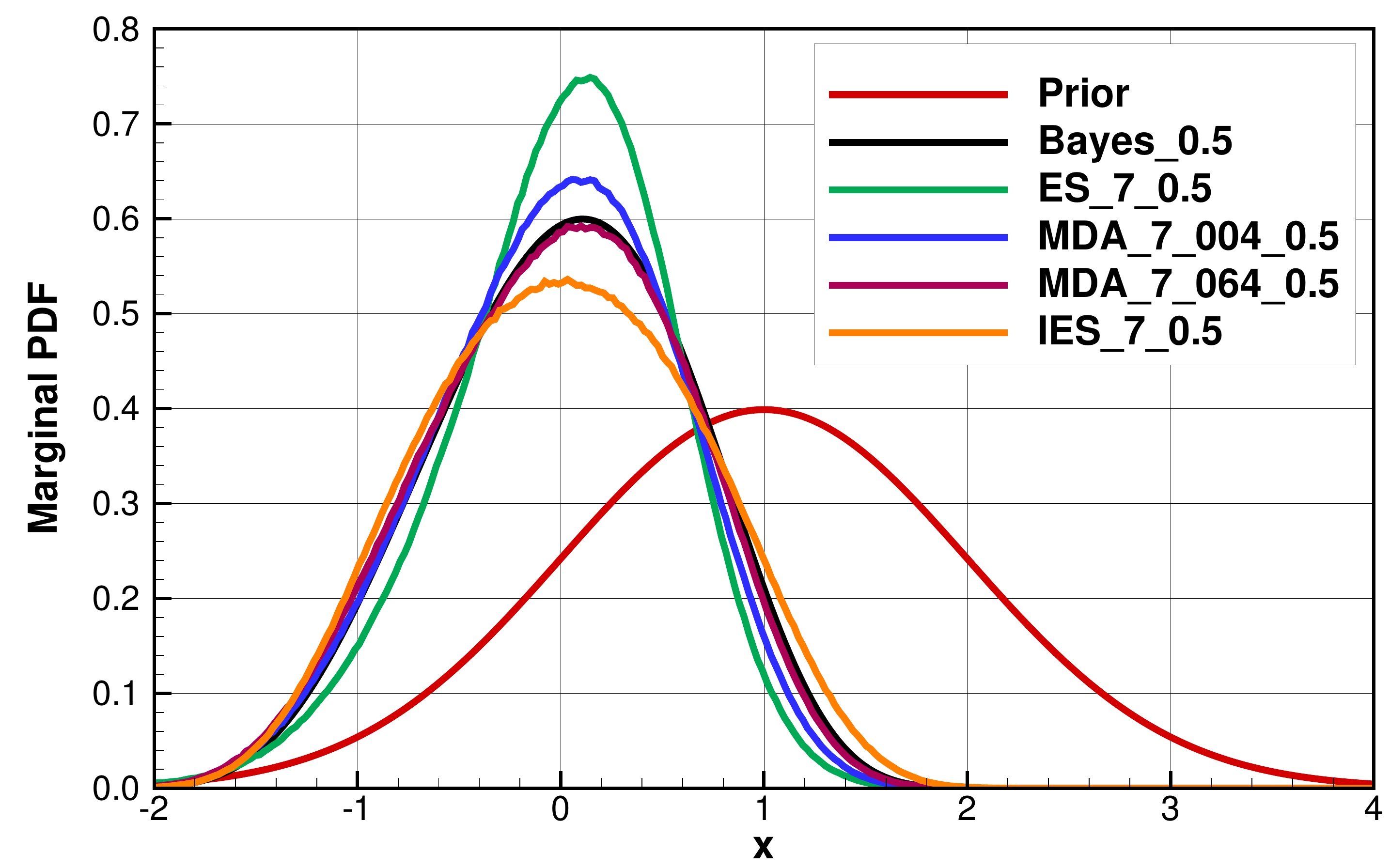}&
\includegraphics[width=0.5\textwidth,trim= 0 0 0 0]{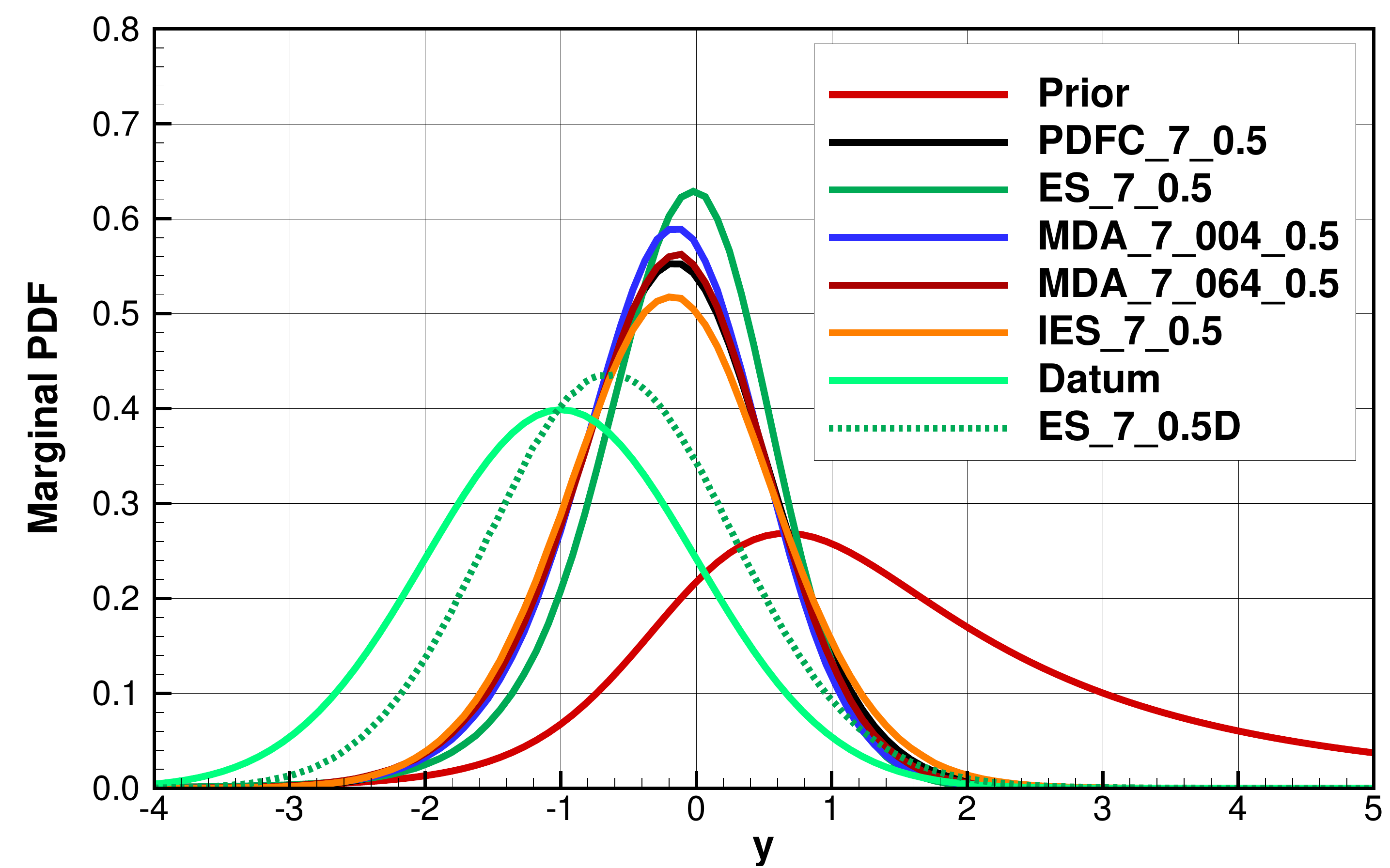}
\end{tabular}
\end{center}
\caption{The plots show the marginal pdfs for $x$ in the left column and the marginal pdfs for $y$ in the right column for the nonlinear case corresponding to Figure~\protect\ref{fig:pdfN}.
The upper row is the case with zero model error, while the lower row shows the result when we include a model error with standard deviation equal to 0.5. In all the plots, the
legends, e.g., MDA$\_7\_004\_0.5$ denote ESMDA with $10^7$ ensemble members, 4 MDA steps, and a model error with variance equal to 0.5.  The legends ES\_7\_0.0D and  ES\_7\_0.5D refers to an ES case where
the prediction $y_j^\rma$ is updated directly by Eq.~(\protect\ref{eq:yupdatege}).
 \label{fig:margN}}
\end{figure*}

\begin{figure}[t]
\begin{center}
\tabcolsep=0.0pt
\begin{tabular}{c}
\includegraphics[width=1.0\columnwidth,trim= 0 0 0 0]{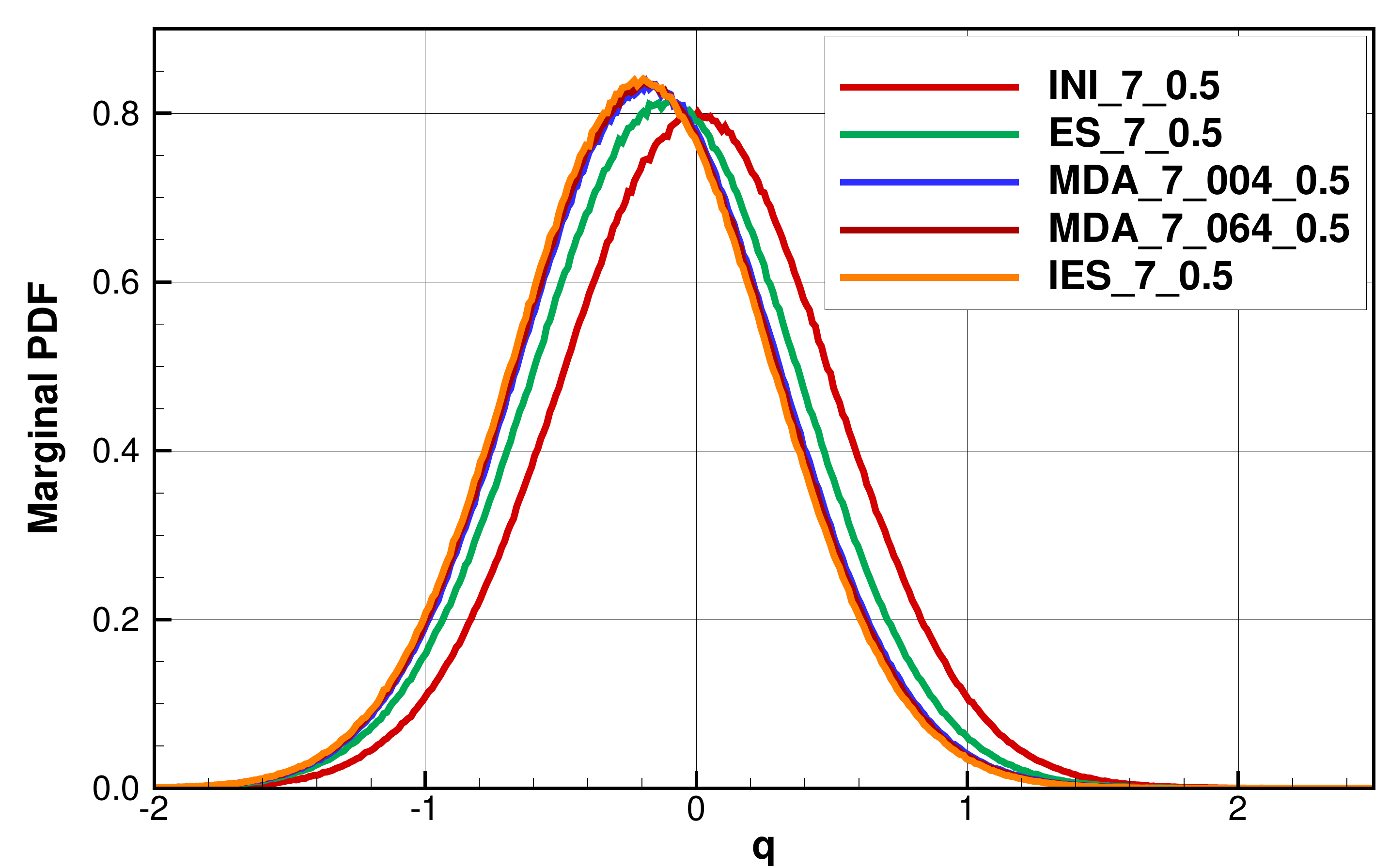}
\end{tabular}
\end{center}
\caption{Model error distributions for the nonlinear case.
 \label{fig:margQN}}
\end{figure}

\begin{figure}[t]
\begin{center}
\tabcolsep=0.0pt
\begin{tabular}{c}
\includegraphics[width=1.0\columnwidth,trim= 0 0 0 0]{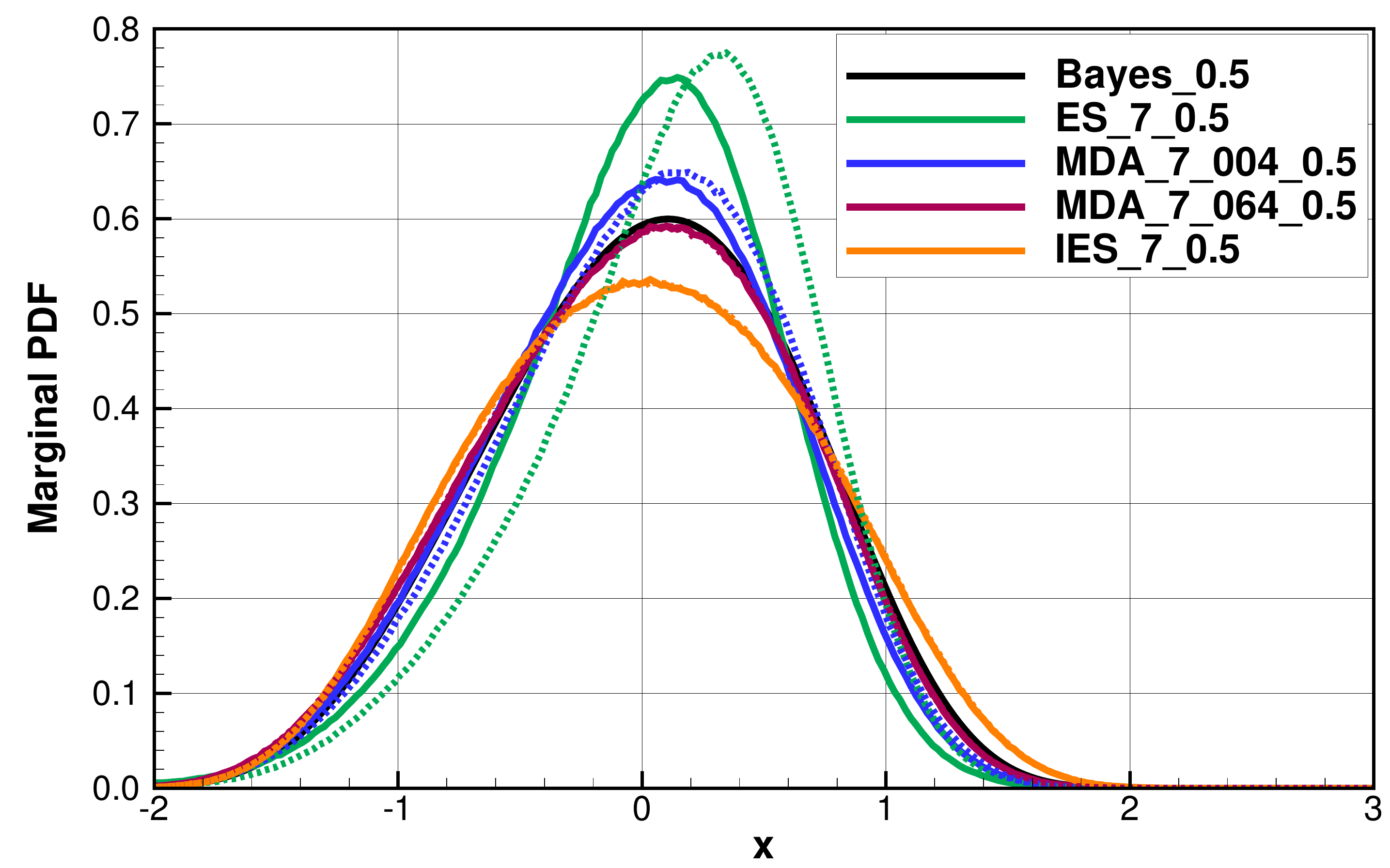}
\end{tabular}
\end{center}
\caption{The plot compares solutions  including (solid lines) and excluding (dashed lines) the projection in Eq.~(\protect\ref{eq:defCyy}).
 \label{fig:projx}}
\end{figure}

\section{Examples\label{sec:ex}}
To verify the new algorithms, we will use the scalar example from \citet{eve18b}.
The example resembles the use of conditioning methods in history matching, i.e., there is a parameter $x$ that serves as
an input to a forward model to predict $y=g(x,q)$.  We assume an initial state $x$ and a prediction $y$, given by the model
\begin{equation}
\begin{split}
 y &= g(x,q)  \\
   &= x (1+\beta x^2) +  q .
\end{split}
\label{eq:scalarmodel}
\end{equation}
Here $\beta$ is a parameter that determines the nonlinearity of the model. In the current example, we have used $\beta=0.0$ for the linear cases and $\beta=0.2$ for the nonlinear cases. Clearly, in this case the
model error is additive to make the linear case completely linear. If we have a product of $x$ and $q$, the problem becomes nonlinear, and we can not test if the different methods converge to the same solution in the
entirely linear case where the convergence partly constitutes our proof of consistency.

The model error $q$ is a random variable sampled from  $\calN(0,C_{qq})$ with $C_{qq}=0.0$ in the case with no model errors and $C_{qq}=0.25$ in the case including model errors.

In all the four cases we sample the prior ensemble for $x$ from a Gaussian distribution with mean $x^\rmf=1$ and variance $C_{xx}=1$ and we sample the perturbed observations of $y$ from a Gaussian 
error distribution with mean $d=-1$ and variance $C_{dd}=1$.
Thus, in the current example, $x$ represents the initial state or the model parameter, while $y$ is the predicted observation. The goal is to estimate $x$ given a measurement of $y$ and then to recompute the correct
prediction of $y$ subject to model errors consistently with Bayes theorem.

In this example, we use a sufficiently large number of samples, i.e., $10^7$, to generate accurate estimates of the probability
density functions and this allows us to work directly with the pdfs and to examine the converged solutions of the methods.

\subsection{Results from the linear case}
In Figs.~\ref{fig:pdfL} and \ref{fig:margL} we show the results from the linear cases without and with model errors.
In Fig.~\ref{fig:pdfL} we plot the joint pdfs for the prior and the updated solutions, and in Fig.~\ref{fig:margL} we plot the corresponding marginal pdfs.

The joint pdfs illustrate the effect of including stochastic model errors. Without model errors, there is a unique mapping from $x$ to $y$, and the pdf is zero except along the
curve (or line in the linear case) defined by the model function $y=g(x)$.  The prior joint pdf has a maximum value located at $(x,y)=(1,1)$ while the posterior joint pdf has shifted the maximum value to
$(x,y)=(0,0)$ for all the methods.
When we introduce the model errors, we notice that we obtain a stronger update in $y$ and weaker update in $x$, than in the case without model errors. Still, we observe that all the smoother methods give a result that
is identical to the Bayesian update.
We can better visualize these results when we examine the marginal pdfs plotted in Fig.~\ref{fig:margL}.

In the case without model errors, we see that the prediction pdf for $y$ and the measurement pdf have the same variance and only differ in the value of the means. The measurement is at $y=-1$ while the mean prediction is
located at $y=1$. The update from ES, ESMDA, and IES, exactly matches the Bayesian update in this case and is centered between the measurement and prediction pdf as we would expect. 

When we include model errors, the
effect is that the prediction gets a higher variance, although the mean is the same (in this particular case). Due to the higher variance, we give more
weight to the measurement in the update and the update for $y$ is stronger than in the case without model errors.
On the other hand, the update for $x$ is weaker in this case, since the addition of model errors reduces the correlation between the predicted measurement $y$ and the prior $x$.  

So, how can the update for $y$ be shifted
towards the observation in this case? Afterall, we compute $y$ as a prediction from $x$. Here the inclusion of the model errors in the inversion plays a vital role. We simultaneously update the ensemble for $x$ and the
estimate of the model errors $q$.  In Fig.~\ref{fig:margQL} we see how we shift the model errors towards negative values.
Thus, when we integrate the model forward from the updated $x$, the forcing from $q$ compensates
for the weaker update of $x$ and also the additional shift of $y$ towards the measured value.

This example illustrates how model errors impact the updates of $x$ and $y$ as well as how we also need to include the model errors as a parameter in the estimation and then use it in 
the prediction to obtain the correct estimate of $y$. Finally, we also demonstrate that in the linear case with and without model errors, ES, IES, and ESMDA, all converge to the correct Bayesian solution.

\subsection{Results from the nonlinear case}
In Figs.~\ref{fig:pdfN} and \ref{fig:margN} we show the results from the nonlinear cases with and without model errors, where
Fig.~\ref{fig:pdfN} plots the joint pdfs for the prior and the updated solutions, and in Fig.~\ref{fig:margN} we plot the corresponding marginal pdfs.

From the joint pdfs, we notice that the various smoother methods give different results both with and without the inclusion of model errors, although the general shape and locations of the
pdfs are reasonably consistent with the theoretical solution as given by Bayes theorem. 

We get a clearer picture from the marginal pdfs in Fig.~\ref{fig:margN}.
As for the linear case, we get a weaker update of $x$ and a stronger update of $y$. We also notice that the introduction of model errors is handled well by the iterative methods, and the 
results are somewhat better and more consistent with the theoretical solution than in the case without model errors.  ES is still the poorest estimator, and the iterative smoothers provide a significant improvement in the
estimate, also in the case including model errors. We show the corresponding updates of the model error $q$ in Fig.~\ref{fig:margQN} and the different smoother methods all give slightly different results.

The dashed green line in the plots for $y$ in Fig.~\ref{fig:margN} is the direct ES update of $y$ using the predicted ensemble for $y$ and the measurement. It is clear that the update of $x$ followed by an
integration of the model to obtain $y$ gives a better result than a direct update of $y$. Furthermore, the additional use of iterations improves the estimate of $y$ even further. This result is the motivation
for introducing IEnKF in sequential data assimilation \citep{sak12a,sak18a} and also the iterative smoother by \citet{boc13a,boc14a}. In history matching, we are primarely interested in estimating $\bmx$, and the
prediction $\bmy$ is just the result given the model parameters in $\bmx$. But also in history matching, the ultimate value comes from accurate predictions of $\bmy$.

The impact of using Eq.~(\ref{eq:defCyy}) for evaluating $\wt{\bmC}_{yy}$ in the update schemes is illustrated in Fig.~\ref{fig:projx} where we show results including and excluding the projection.
The impact is most pronounced when using ES and ESMDA with few update steps where the use of $\ol{\bmC}_{yy}$ instead of $\wt{\bmC}_{yy}$ severely impacts the computation of the long linear update steps. 
In IES we must include the projection to ensure that the gradients defined in Eqs.~(\ref{eq:DeltazB}) and (\ref{eq:DeltazB2}) are identical but in the current case the relative difference in the estimated mean when including or
excluding the projection is only around one percent.

\section{Including model errors in history matching}
The need for including model errors in iterative ensemble smoothers became apparent while working with the paper \citep{eve18a}, which considered the conditioning of reservoir models on production-rate data.
Typically, in history matching, we assign errors to the rate data used in the conditioning step, while we neglect these errors when the same data are used to force the reservoir simulation model during the historical
simulation.

The errors in rate data are considered as the dominant model errors in a reservoir simulation model when we exclude errors in the model parameters that we estimate during the history matching.
Also, \citet{eve18a} pointed out that there are strong time correlations in the errors in rate data due to the rate allocation procedure used. When we include a stochastic model forcing using time-correlated errors, we
will experience a  significantly stronger impact than when the errors are white in time \citep[see Chap.~12 in][]{eve09book}.

The functional form $\bmy=\bmg(\bmx, \bmq)$ can represent the prediction of the produced rates (that we observe) from a reservoir simulation model. Note that using ensemble methods, we do not need to
explicitly construct the functional form $\bmy=\bmg(\bmx, \bmq)$ since we represent the gradients using ensemble covariances. We only need access to a numerical reservoir simulation model.
The inputs $\bmx$ then contain all the uncertain parameters of the model, such as, e.g., porosity and permeability fields, various transmissibilities, and even parameters defining the reservoir structure.
The model errors can be the errors in the rates used to force the model in the historical period. These errors are not additive since the model prediction depends nonlinearly on the specified rates.

If we associate the dominant model errors with the rates used to force the simulation model, then the size of the vector of model errors $\bmq$ is equal to the number of rate data used to force the model. A typical number
of data for a well can be 12 data points per year, i.e., if we force the model using monthly reservoir-volume rates.

The prior error statistics for the rate data used to force the model should be the same as is used for the rate data in the update step. Thus, we can simulate a prior ensemble of time series with mean zero, a
specified variogram in time, and a specified variance, to represent the model errors. These time series are then defining the vectors $\bmq_j$ of model errors for each realization. An ensemble model-error covariance
$\ol{\bmC}_{qq}$ is then defined by the ensemble $\bmq_j$. 

The conventional procedure of deriving the production rates from rate-allocation tables, which we only update in connection with separator tests, often several years apart, means that the errors in the rate data will be
nearly perfectly correlated in between each separator test.  Note also that the inclusion of time correlations significantly reduces the degrees of freedom in the model errors and simplifies the estimation of the model
errors in the conditioning step.

The expected impact of including model errors is first a larger and more realistic spread of the prior ensemble.  Second, we will obtain a weaker and more correct update of the reservoir parameters in conditioning 
step. Furthermore, the posterior ensemble will give a more accurate and consistent prediction at the end of the history matching period since the posterior realizations are forced by updated and improved estimated rates.
Thus, we have an improved basis for making future predictions. We also expect that the information contained in the improved estimates of rates, in some cases may be used to correct for biases in the rates used in future
predictions.

\section{Summary}
In this paper, we have given consistent formulations of iterative ensemble smoothers when we include model errors. We demonstrate the consistency by showing that the ensemble smoothers all converge to the Bayesian
posterior in the linear case. The main result is that the model errors need to be treated as another set of unknown parameters and estimated in the same way as the input parameters to the model.  The proposed approach
allows for the inclusion of general nonlinear errors that can be both red and white in time. Thus, we can apply the methods for
history matching of reservoir models forced with uncertain rate data having time-correlated errors, as well as an iterative EnKF in sequential data assimilation with general model-error terms.

We demonstrate that the inclusion of model errors leads to a weaker update of the input parameters to the model, but a stronger update of the measured model prediction. Vice verse, the negligence of model errors
that should be present, will lead to a too substantial update of the model input parameters with an associated underestimated uncertainty and also a too weak update of the prediction.

Thus, the results open for a more consistent solution of the history-matching problem, given that significant model errors are neglected in all previous history-matching applications with iterative ensemble smoothers.

We also briefly discuss an inconsistency of the linearizations in the analysis scheme that appear for nonlinear operators and when the state dimension is less than the ensemble size minus one, and we show that we must
include an additional projection of the predicted measurements for consistency in the derivation of the final update equations.


\appendix
\section*{Acknowledgement}
This work was supported by the Research Council of Norway and
the companies Aker–BP, DEA, ENI, Shell, Petrobras, Equinor, and
VNG, through the Petromaks--2 project DIGIRES. Also, the work has benefited from the interaction and collaborations with the Nordforsk Nordic center of excellence in data assimilation, EMBLA. In-depth discussions with
Patrick Raanes regarding the linear-regression derivation have helped to improve the manuscript, and the author is also grateful for comments by Geir N{\ae}vdal and Alberto Carrassi on early versions of
the manuscript. Constructive comments by three anonymous reviewers lead to the inclusion of a section on how to practically account for
model errors in reservoir history matching.


\end{document}